\pdfoutput=1

\documentclass[twocolumn,showpacs,showkeys,amsmath,amssymb,superscriptaddress,nofootinbib,floatfix]{revtex4-1} 

\usepackage{epsfig}

\newcommand{\nucBea}{\ensuremath{{}^{7}{\rm Be}+{}^{208}{\rm Pb}}}
\newcommand{\nucBeb}{\ensuremath{{}^{9}{\rm Be}+{}^{208}{\rm Pb}}}
\newcommand{\nucBec}{\ensuremath{{}^{7}{\rm Be}}}

\newcommand{\nucBee}{\ensuremath{{}^{9}{\rm Be}}}
\newcommand{\nucCa}{\ensuremath{{}^{12}{\rm C}+{}^{208}{\rm Pb}}}
\newcommand{\nucCb}{\ensuremath{{}^{12}{\rm C}}}
\newcommand{\nucOa}{\ensuremath{{}^{16}{\rm O}+{}^{208}{\rm Pb}}}
\newcommand{\nucOb}{\ensuremath{{}^{16}{\rm O}}}

\begin{document} 
\hbadness=10000

\title{Signatures of $\alpha$ clustering in ultra-relativistic collisions with light nuclei}

\author{Maciej Rybczy\'nski}
\email{Maciej.Rybczynski@ujk.edu.pl} 
\affiliation{Institute of Physics, Jan Kochanowski University, 25-406 Kielce, Poland}  

\author{Milena Piotrowska}
\email{milena.soltysiak@op.pl} 
\affiliation{Institute of Physics, Jan Kochanowski University, 25-406 Kielce, Poland}

\author{Wojciech Broniowski}
\email{Wojciech.Broniowski@ifj.edu.pl}
\affiliation{Institute of Physics, Jan Kochanowski University, 25-406 Kielce, Poland}
\affiliation{H. Niewodnicza\'nski Institute of Nuclear Physics PAN, 31-342 Cracow, Poland}

%\date{27 October 2017}  

\begin{abstract}
We explore possible observable signatures of $\alpha$ clustering of light nuclei in ultra-relativistic nuclear collisions 
involving ${}^{7,9}$Be, ${}^{12}$C, and ${}^{16}$O. The clustering leads to specific spatial correlations of the nucleon 
distributions in the ground state, which are manifest in the earliest stage of the  ultra-high energy reaction. The formed initial state of the fireball 
is sensitive to these correlations, and the effect influences, after the collective evolution of the system, the hadron production
in the final stage. Specifically, we study effects on the 
harmonic flow in collisions of light clustered nuclei with a heavy target (${}^{208}$Pb), showing that measures of the elliptic 
flow are sensitive to clusterization in ${}^{7,9}$Be, whereas triangular flow is sensitive to clusterization in  ${}^{12}$C and ${}^{16}$O. 
Specific predictions are made for model collisions at the CERN SPS energies.
In another exploratory development we also examine the proton-beryllium collisions, where the $3/2^-$ ground state of ${}^{7,9}$Be nucleus 
is polarized by an external magnetic field. Clusterization leads to multiplicity distributions of participant nucleons which depend on the orientation of 
the polarization with respect to the collision axis, as well as on the magnetic number of the state. The obtained 
effects on multiplicities reach a factor of a few for collisions with a large number of participant nucleons.
\end{abstract}



\maketitle

\section{Introduction}
\label{sec:intro}

The structure of nuclei involving $\alpha$ clusters continues to be a subject of very active studies (see \cite{Freer:2017gip} 
for a recent review, \cite{brink2008history} for a historical perspective, \cite{Arriola:2014lxa} for a discussion of clustering mass formulas and form factors as 
manifestations of the geometric structure,
and \cite{brink1965alpha,freer2007clustered,ikeda2010clusters,Okolowicz:2012kv,beck2012clusters,Beck:2014fja} for additional 
information), exploring the ideas dating back to Gamow's original clusterization proposal~\cite{gamow1931constitution} 
with modern theoretical~\cite{Funaki:2006gt,Chernykh:2007zz,KanadaEn'yo:2006ze} and
computational~\cite{PhysRevLett.109.052501,Barrett:2013nh,Epelbaum:2012qn,Pieper:2002ne,Wiringa:2013ala,Lonardoni:2017egu} methods, as well as with 
anticipated new experimental prospects~\cite{Yamaguchi:2012sz,Zarubin,Fritsch:2017rxc,Guo:2017tco}. 

\begin{figure}[tb]
\centering
\vspace{-2mm}
\includegraphics[angle=0,width=0.40 \textwidth]{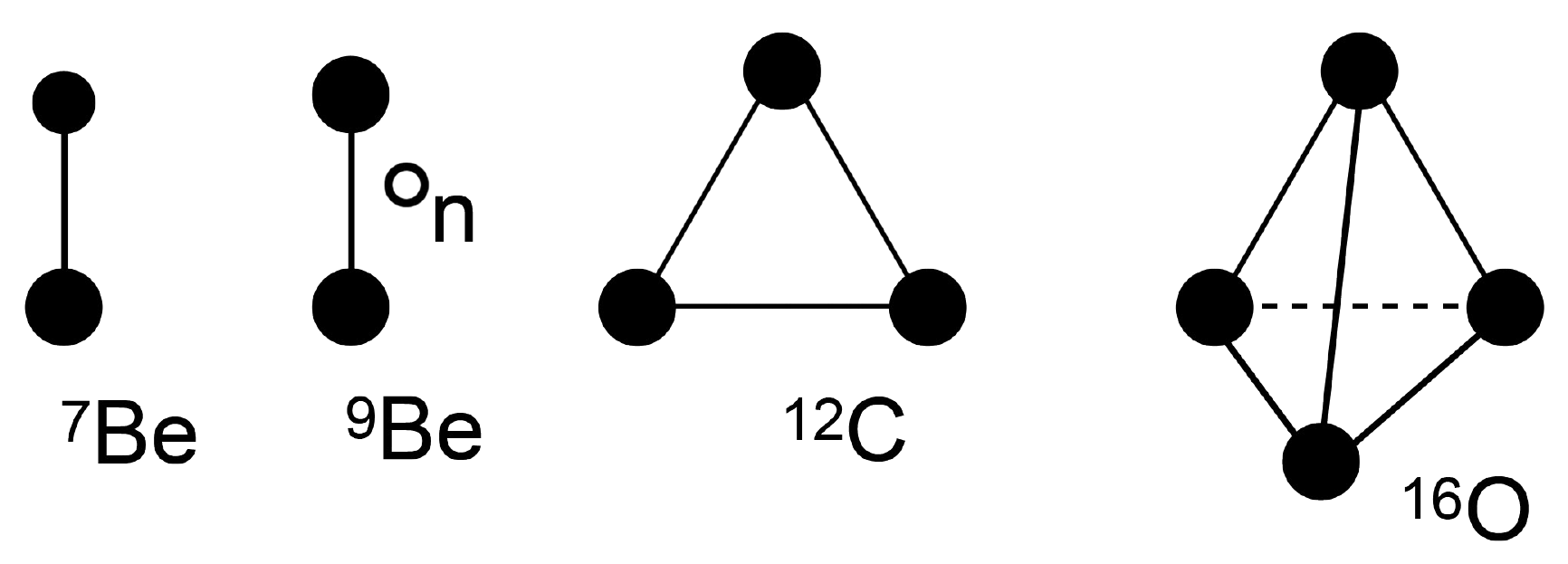}
\vspace{-4mm}
\caption{Schematic view of the cluster structure of light nuclei. The dark blobs indicate $\alpha$ clusters (in the case of ${}^7$Be, also the ${}^3$He cluster). The 
additional dot in ${}^9$Be indicates the extra neutron.
\label{fig:structure}}
\end{figure}

A few years ago a possible approach of investigating $\alpha$ clustering in light nuclei via studies of ultra-relativistic nuclear 
collisions was proposed in Ref.~\cite{Broniowski:2013dia} and explored in further 
detail for the ${}^{12}$C nucleus in~Ref.~\cite{Bozek:2014cva}. Quite remarkably, the experimental application of the method could reveal information on 
the {\em ground state} of a light clustered nucleus, i.e. on the lowest possible energy state, via the highest-energy nuclear collisions, such as those carried out at 
ultra-relativistic accelerators: the CERN Super Proton Synchrotron (SPS), BNL Relativistic Heavy-Ion Collider (RHIC), or the CERN Large Hadron Collider (LHC).
In the first part of this paper we extend the results of Refs.~\cite{Broniowski:2013dia,Bozek:2014cva} obtained for  ${}^{12}$C  to other light nuclei, 
namely ${}^{7}$Be, ${}^{9}$Be, and ${}^{16}$O, which are believed to have a prominent cluster structure in their ground states, see Fig.~\ref{fig:structure}.

We recall the basic concepts of Refs.~\cite{Broniowski:2013dia,Bozek:2014cva}: Spatial correlations in the ground state of a light nucleus, such as the 
presence of clusters, lead to an intrinsic deformation. When colliding with a heavy nucleus (${}^{208}$Pb,  ${}^{197}$Au) at a very high energy, 
where due to the Lorentz contraction the collision time is much shorter than any characteristic nuclear time scale, 
a reduction of the wave function occurs and a correlated spatial distribution of participant nucleons is formed.
This, via individual nucleon-nucleon collisions between the colliding nuclei in the applied Glauber 
picture~\cite{Glauber:1959aa,Czyz:1968zop,Bialas:1976ed}, leads to an initial distribution of entropy in the 
transverse plane, whose {\em eccentricity} reflects the deformation of the ground-state due to correlations. 
In short, the deformed intrinsic shape of the light nucleus, when hitting a ``wall'' of a heavy target, yields a 
deformed fireball in the transverse plane.

\begin{figure}[tb]
\centering
\vspace{-2mm}
\includegraphics[angle=0,width=0.45 \textwidth]{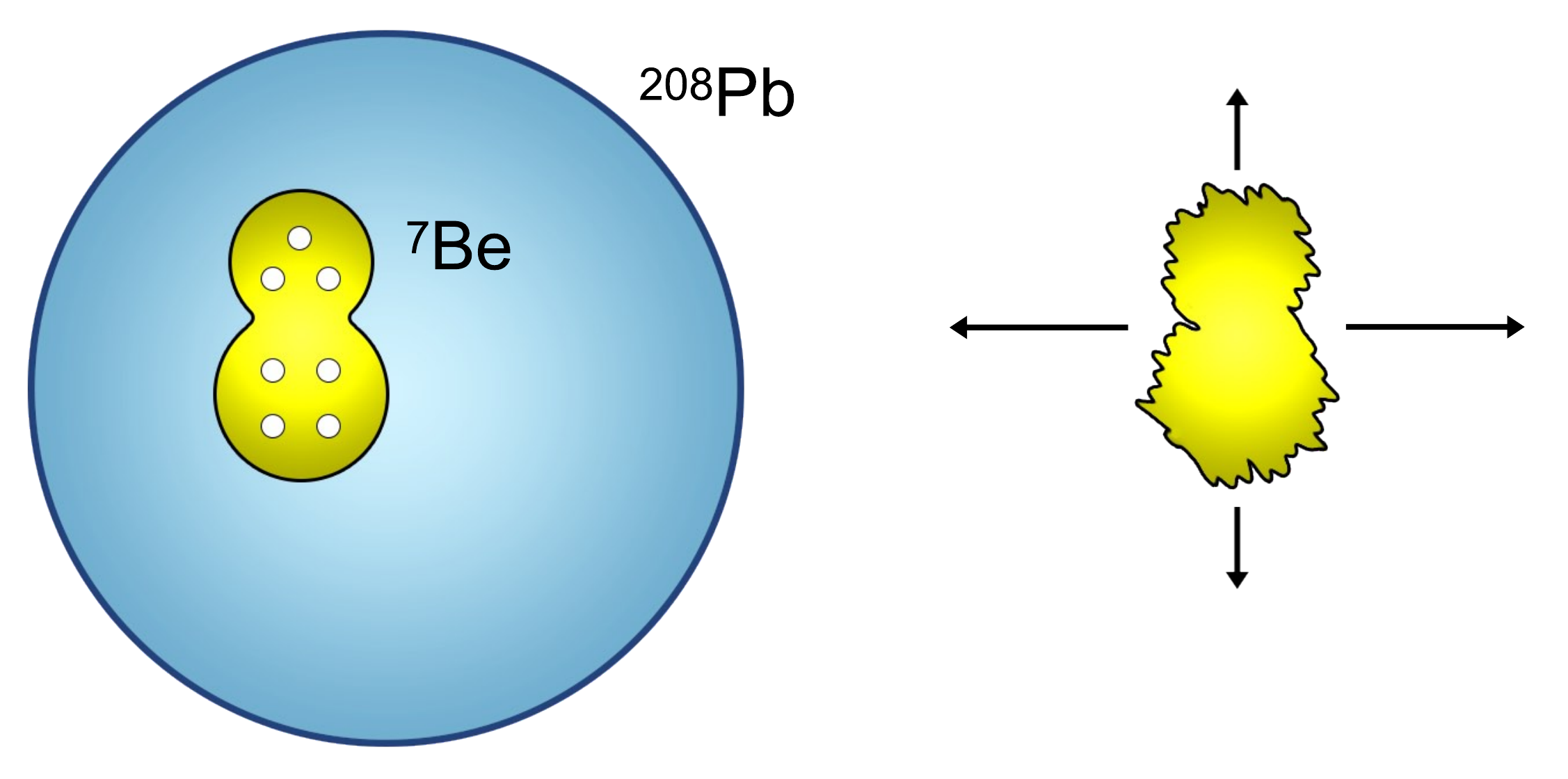}
\vspace{-1mm}
\caption{Cartoon of ultra-relativistic $^{7,9}$Be+$^{208}$Pb collisions.
The clustered beryllium creates a fireball whose initial transverse shape reflects the deformed intrinsic shape of the projectile (left panel). 
Subsequent collective evolution leads to faster expansion along the direction perpendicular to the symmetry axis of the beryllium, and slower expansion 
along this axis, as indicated by the arrows (right panel). The effect generates specific signatures in the harmonic flow patterns in spectra of the produced hadrons in the final state.
\label{fig:concept}}
\end{figure}

As an example, 
if the intrinsic state of the ${}^{12}$C nucleus is a triangle made of three $\alpha$ particles, then the shape of the initial 
fireball in the transverse plane reflects this triangular geometry. Next, the {\em shape-flow transmutation} mechanism (cf. Fig.~\ref{fig:concept}), 
a key geometric concept in the phenomenology of ultra-relativistic heavy-ion collisions~\cite{Ollitrault:1992bk}, generates a large collective 
triangular flow through the dynamics in the later stages of the evolution, modeled via hydrodynamics (for recent reviews see~\cite{Heinz:2013th,Gale:2013da,Jeon:2016uym}), 
or transport~\cite{Lin:2004en}.  As a result, one observes the azimuthal asymmetry of the transverse momentum distributions
of produced hadrons. Similarly, the dumbbell 
intrinsic shape of the ground states of the ${}^{7,9}$Be nuclei, which occurs when  these nuclei are clustered, leads to a large elliptic flow. 

We remark that the methodology applied in Refs.~\cite{Broniowski:2013dia,Bozek:2014cva} and in the present work, 
was used successfully to describe harmonic flow in d+Au collisions~\cite{Bozek:2011if} (small dumbbells)  
and in $^3$He+Au collisions~\cite{Nagle:2013lja,Bozek:2014cya} (small triangles), 
and the predictions later experimentally confirmed in~\cite{Adare:2013piz,Adare:2015ctn}.

As the positions of the nucleons in the colliding nuclei fluctuate, being distributed according to their wave functions, the initial eccentricity, 
and in consequence the harmonic flow, always receives an additional contribution from 
these random fluctuations~\cite{Miller:2003kd,Alver:2006wh,Voloshin:2006gz,Broniowski:2007ft,Hama:2007dq,Luzum:2011mm,Bhalerao:2014xra} (the shape fluctuations are indicated 
with a warped surface of the fireball in Fig.~\ref{fig:concept}). For that reason the applied measures of the harmonic flow should be able to 
discriminate between these two components.

To a good  approximation, the measured harmonic flow coefficients $v_n$ in the spectra of produced hadrons are linear in the corresponding initial eccentricities 
$\epsilon_n$ (see, e.g., ~\cite{Gardim:2011xv,Niemi:2012aj,Bzdak:2013rya}).  This allows for a construction of flow measures given 
in Sec.~\ref{sec:signatures}, which are independent the of details of the dynamics of the later stages of the collision, and thus carry 
information pertaining to the initial eccentricities. We describe such measures in Sect.~\ref{sec:signatures}.
We note that another measure, involving the ratio of the triangular and elliptic flow coefficients, 
has been recently proposed in Ref.~\cite{Zhang:2017xda} 
for the case of ${}^{12}$C, and tested within the AMPT~\cite{Lin:2004en} transport model.

To have realistic nuclear distributions with clusters, yet simple enough to be implemented in a Monte Carlo simulation,
we apply a procedure explained in Sec.~\ref{sec:making}, where positions of nucleons are determined within clusters of a given size, whereas 
the clusters themselves are arranged in an appropriate shape (for instance, triangular for ${}^{12}$C.  
The parameters, determining the separation distance between the clusters and their sizes,  
are fixed in such a way that the resulting one-body nucleon densities compare well to the state-of-the-art Variational Monte Carlo 
(VMC)~\cite{Wiringa:2013ala,Lonardoni:2017egu} simulations. 
The simulations for clustered nuclei are compared to the base-line case, where no clustering is present.

Our basic findings, presented in Sec.~\ref{sec:signatures}, are that clusterization in light nuclei leads to sizable effects in the harmonic flow 
pattern in collisions with heavy nuclei. The effect is most manifest for the highest-multiplicity collisions, where additional fluctuations 
from the random distribution of nucleons are reduced. For the dumbbell shaped ${}^{7,9}$Be, the measures of 
the elliptic flow are affected, whereas for the triangular ${}^{12}$C and tetrahedral ${}^{16}$O there are significant imprints of clusterization in 
the triangular flow. These effects, when observed experimentally, could be promptly 
used to assess the degree of clusterization in light nuclei.

In the second part of this paper we examine a novel possibility of observing the intrinsic deformation  resulting from clusterization 
 of light nuclei with spin, such as ${}^{7,9}$Be, when these are collided with ultra-relativistic {\em protons}. 
This interesting but exploratory proposal would require a magnetically 
{\em polarized} ${}^{7,9}$Be nuclei, which in the ground state have $J^P=3/2^-$. 

In this case the geometric mechanism is as follows:
When the dumbbell shaped nucleus in $m=1/2$ ground state is polarized along the proton beam direction, there is a much higher chance 
for the proton to collide with more nucleons (as it can pass through both clusters) than in the case where it is polarized perpendicular to the beam axis 
(where it would pass through a single cluster only). Thus more participants are formed in the former case. 
The effect is opposite for the $m=3/2$ state, as explained in Sect.~\ref{sec:pA}.

One could thus investigate the distribution of participant nucleons, $N_w$, for various magnetic numbers $m$ and geometric orientations.
We find from our simulations a factor of two effects for $N_w = 4 $ and an order of magnitude effect for $N_w \ge 6$, when 
comparing the cases of \mbox{$m=3/2$} and \mbox{$m=1/2$}, or changing of the direction of the beam relative 
to the polarization axis. We discuss the mechanism and the relevant issues in Sec.~\ref{sec:pA}.

\section{Nucleon distributions in clustered light nuclei}
\label{sec:making}

To model the collision process in the applied Glauber framework~\cite{Glauber:1959aa,Czyz:1968zop,Bialas:1976ed}, we 
first need the distributions of centers of nucleons in the considered nuclei. We have adopted a simple and practical procedure 
where these distributions are generated randomly in clusters placed at preassigned positions in such a way that 
the one-body density reproduces the shapes obtained from state-of-the-art 
Variational Monte Carlo (VMC)~\cite{Buendia:2004yt,Wiringa:2013ala,Lonardoni:2017egu} studies.

Explicitly, our steps are as follows: We set the positions of clusters according to the geometry of Fig.~\ref{fig:structure}, separating their centers
from each other with the distance $l$. The distribution of the nucleons in each cluster is randomly generated according to the Gaussian function
\begin{eqnarray}
f_i(\vec{r})=A \exp \left (- \frac{3}{2} \, \frac{(\vec{r}-\vec{c_i})^2}{r_c^2} \right ),   
\end{eqnarray}
where $\vec{r}$ is the 3D coordinate of the nucleon, $\vec{c_i}$ is the position of the center of the cluster $i$, and $r_c$ is the rms radius 
of the cluster, which equals $r_\alpha$ or  $r_{{}^3{\rm He}}$ depending on the cluster type. 
We generate the positions of the nucleons in sequence, alternating the number of the cluster: 1, 2,\dots, 1, 2,\dots, until all the nucleons are 
placed.

\begin{table}[tb]
\caption{\label{tab:param} Parameters used in the GLISSANDO simulations to obtain the nuclear distribution: $l$ is the distance between the centers 
of clusters, arranged according to the geometry shown in Fig.~\ref{fig:structure}, $r_\alpha$ is the size of the $\alpha$ cluster, $r_{{}^3{\rm He}}$ 
is the size of the  ${}^3{\rm He}$ cluster in \nucBec, and $r_n$ determines the distribution of the extra neutron in \nucBee.}
\vspace{3mm}
\begin{tabular}{|c|cccc|}
\hline
 Nucleus & $l$ [fm] & $r_\alpha$ [fm] &  $r_{{}^3{\rm He}}$ [fm]  & $r_n$ [fm]
\\
\hline 
\nucBec   & 3.2 & 1.2 & 1.4 & - \\
\nucBee   & 3.6 & 1.1 & -  & 1.9  \\
\nucCb     & 2.8 & 1.1 & - & - \\
\nucOb    & 3.2 & 1.1 & - & - \\
\hline
\end{tabular}
\end{table}

\begin{figure}[b]
\centering
%\vspace{-2mm}
\includegraphics[angle=0,width=0.48\textwidth]{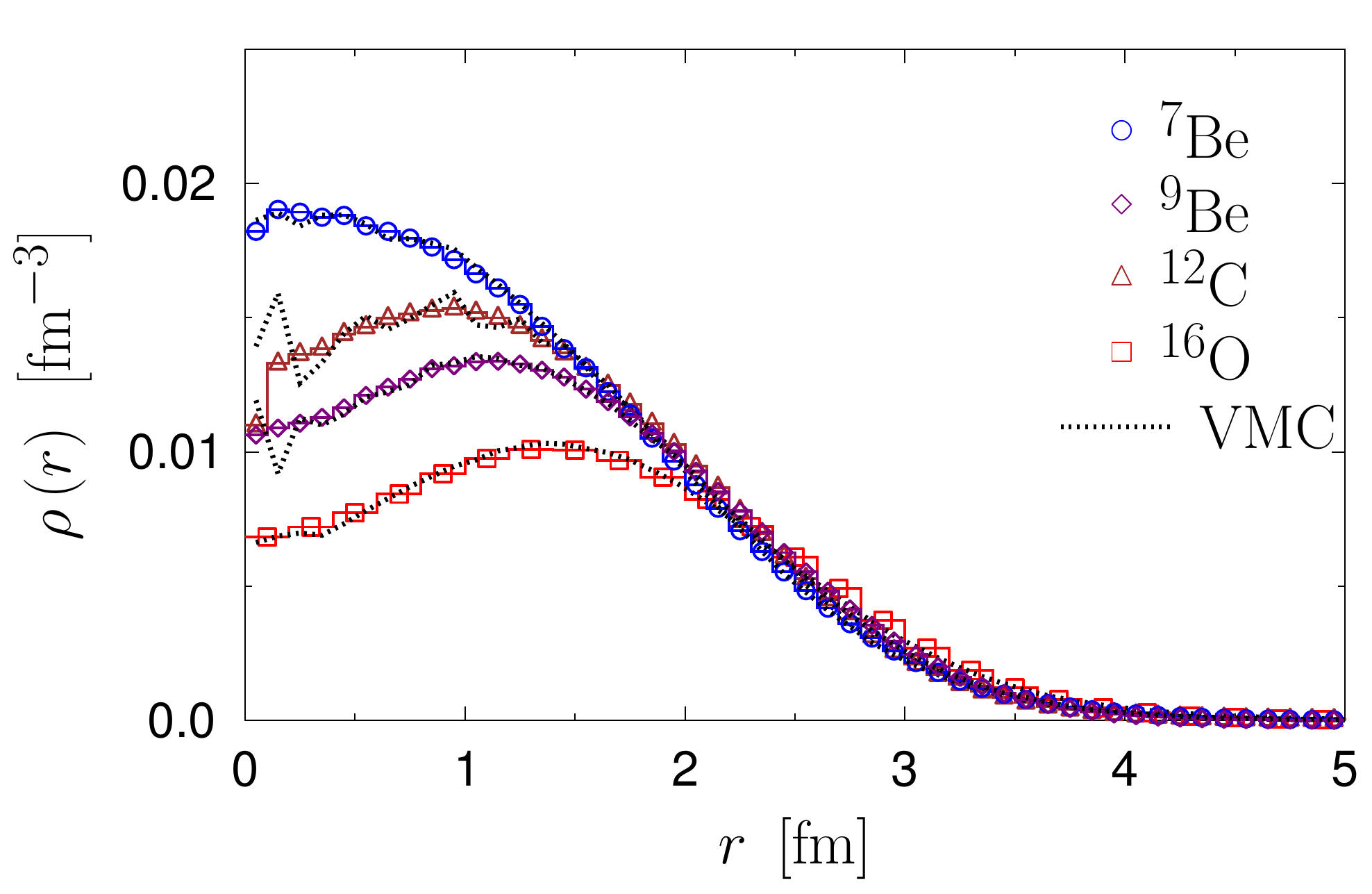}
%\vspace{-8mm}
\caption{(Color online) Nuclear density profiles of the considered light nuclei. The points correspond to our Monte Carlo generation of the nuclear distributions 
in GLISSANDO, with parameters listed of Table~\ref{tab:param} adjusted in such a way that the results from  
Variational Monte Carlo (VMC)~\cite{Wiringa:2013ala,Lonardoni:2017egu}(dashed lines)  are properly reproduced. 
We use the normalization $4\pi \int_0^\infty r^2dr \, \rho(r)=1$. \label{fig:density}}
\end{figure}

For ${}^{9}$Be, we add the extra neutron on top of the two $\alpha$ clusters according to a distribution with a hole in the middle, 
\begin{eqnarray}
f_n(\vec{r})=A' r^2 \,\exp \left (- \frac{3}{2} \frac{r^2}{r_n^2} \right ).
\end{eqnarray}

The short-distance
nucleon-nucleon repulsion is incorporated by precluding the centers of each pair of nucleons to be closer than the expulsion distance of 
0.9~fm, which is a customary prescription~\cite{Broniowski:2010jd} in preparing nuclei for the Glauber model in ultra-relativistic nuclear collisions. 
At the end of the procedure the distributions are shifted such that their center of mass is 
placed at the origin of the coordinate frame.
As a result, we get the Monte Carlo distributions with the built-in cluster correlations.

\begin{figure*}[tb]
\centering
%\vspace{-2mm}
\includegraphics[angle=0,width=0.495 \textwidth]{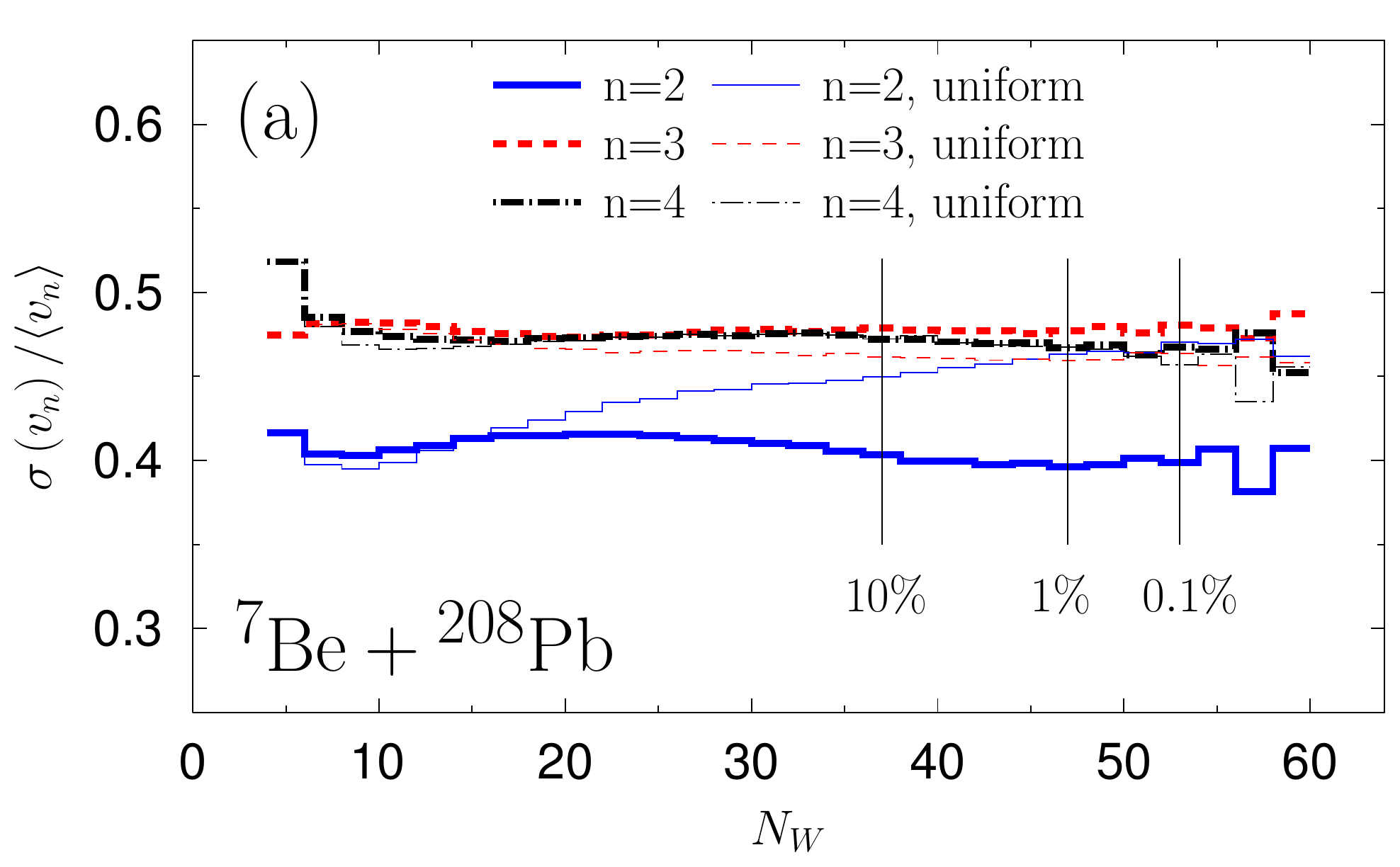}
\includegraphics[angle=0,width=0.495 \textwidth]{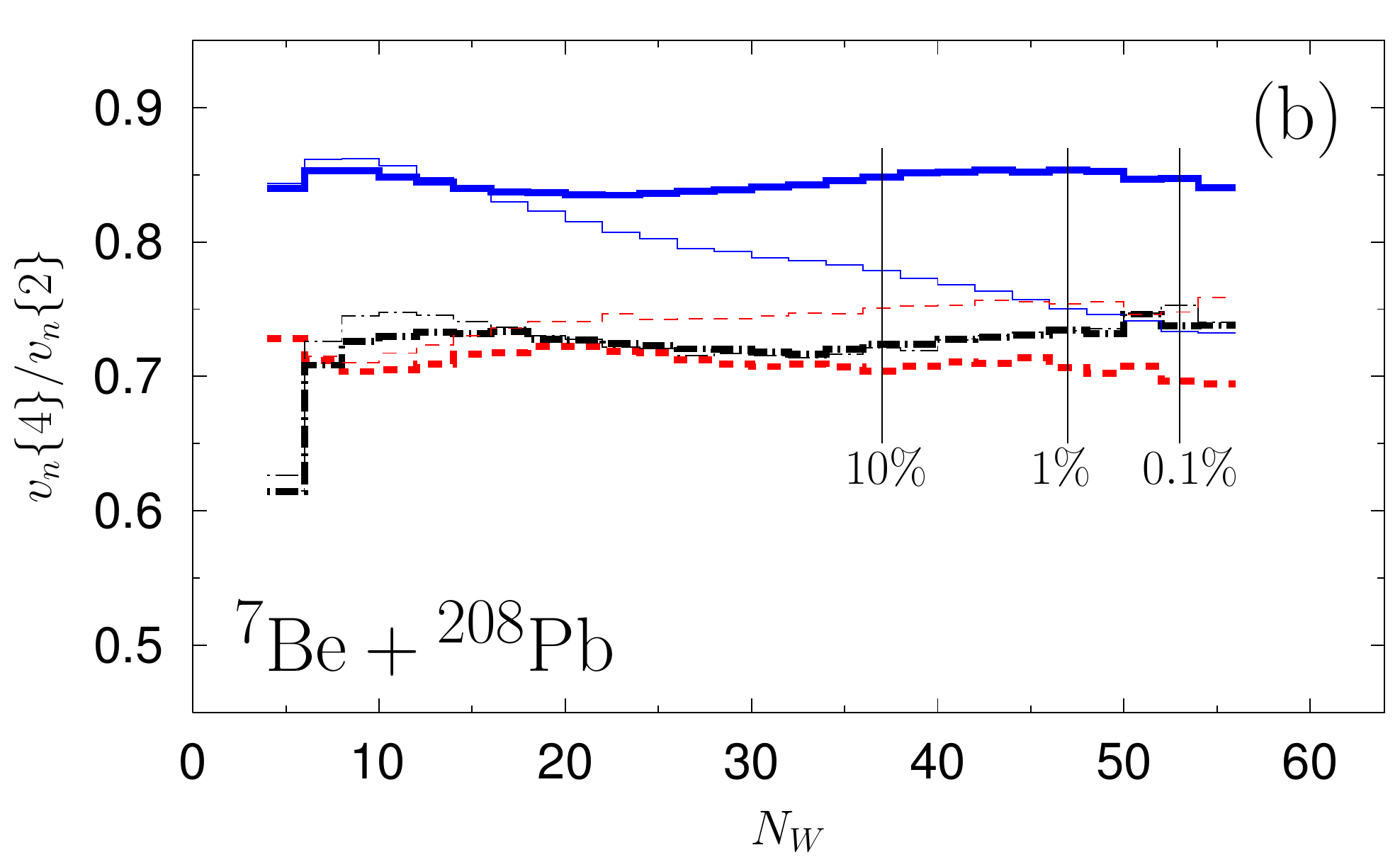}
%\vspace{-8mm}
\caption{(Color online) Scaled standard deviations of rank-$n$ flow coefficients (panel (a)) and ratios of the four-particle to two-particle 
cumulants (panel (b)), plotted as functions of the total number of the wounded nucleons. Clustered nuclei (thick lines) are compared with 
the case where the nucleons are distributed uniformly with the same one-body radial distributions (thin lines). \nucBea\ collisions.
The vertical lines indicate the multiplicity percentiles (centralities) corresponding to the indicated values of $N_w$. 
}
\label{fig:7Be}
\end{figure*}

\begin{figure*}[tb]
\centering
%\vspace{-2mm}
\includegraphics[angle=0,width=0.495 \textwidth]{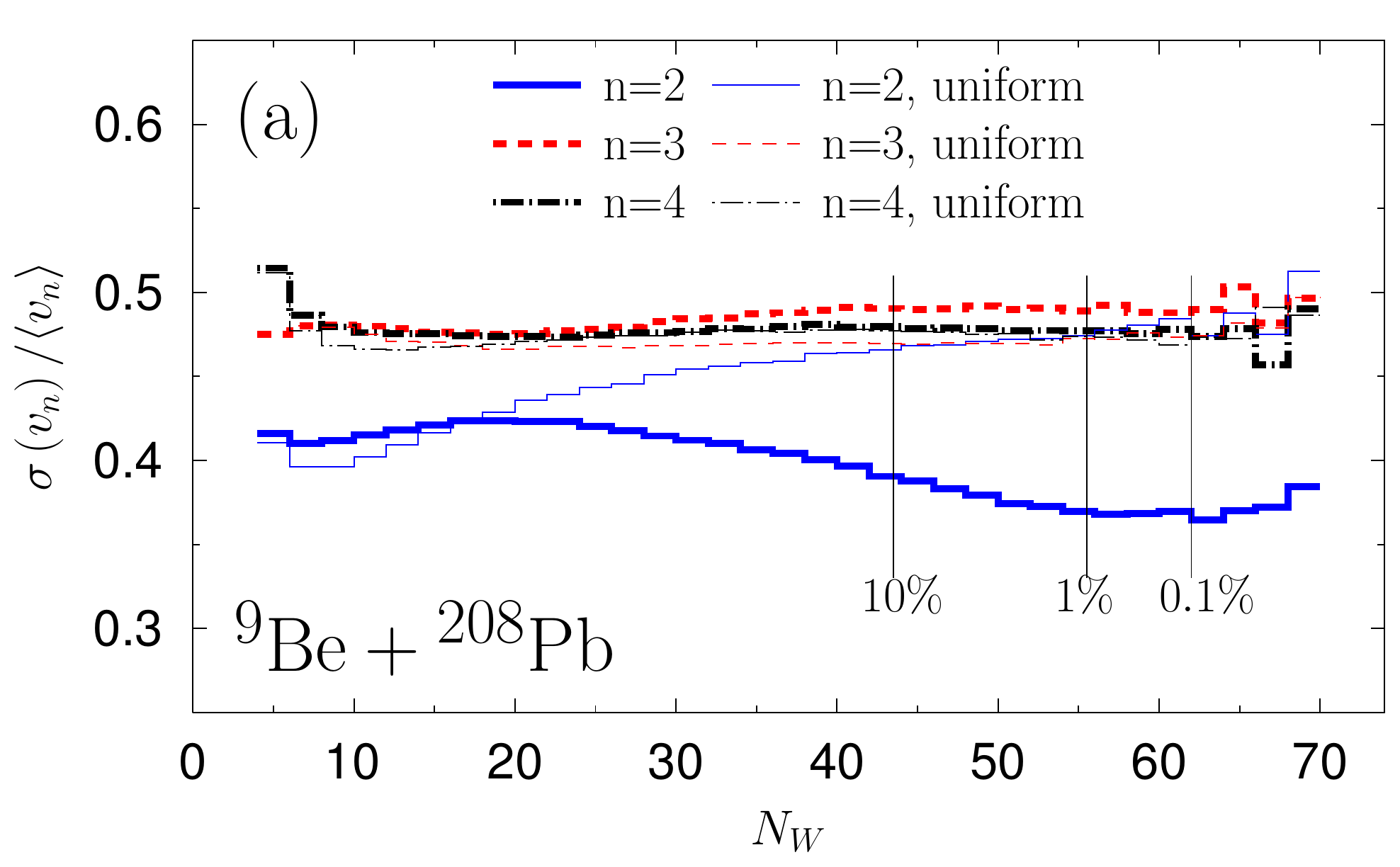}
\includegraphics[angle=0,width=0.495 \textwidth]{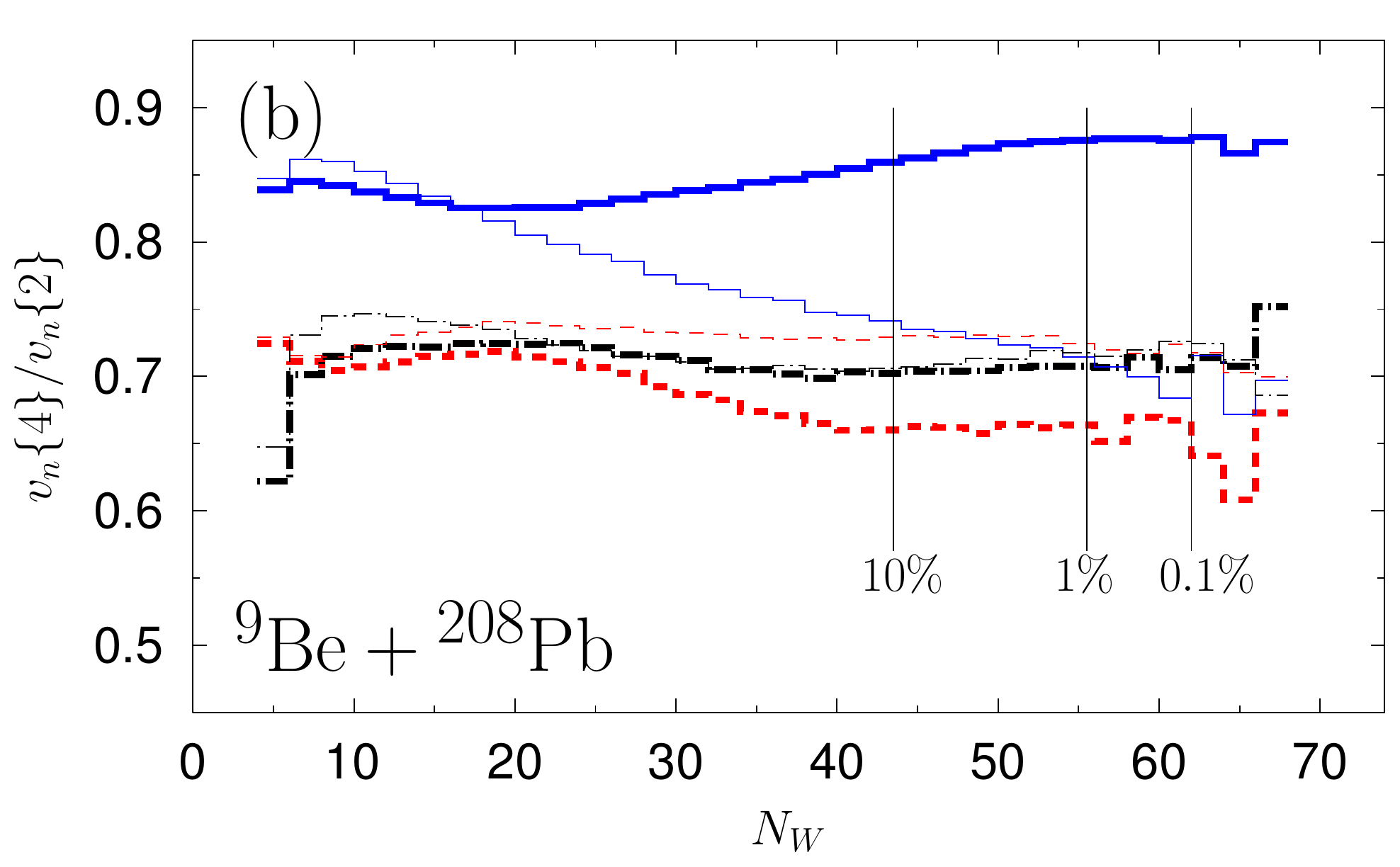}
%\vspace{-8mm}
\caption{(Color online) The same as in Fig.~\ref{fig:7Be} but for \nucBeb\ collisions.}
\label{fig:9Be}
\end{figure*}

\begin{figure*}[tb]
\centering
%\vspace{-2mm}
\includegraphics[angle=0,width=0.495 \textwidth]{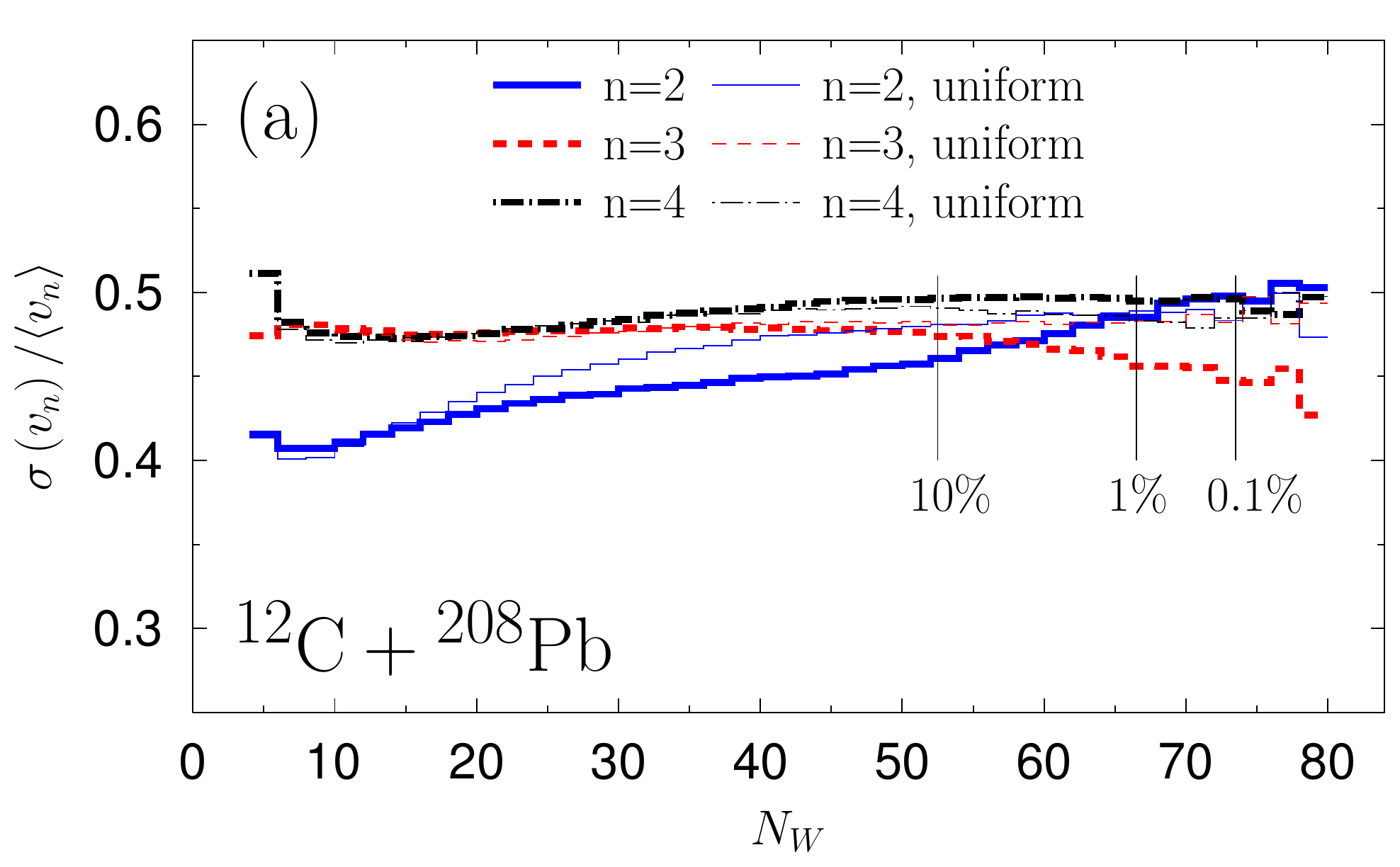}
\includegraphics[angle=0,width=0.495 \textwidth]{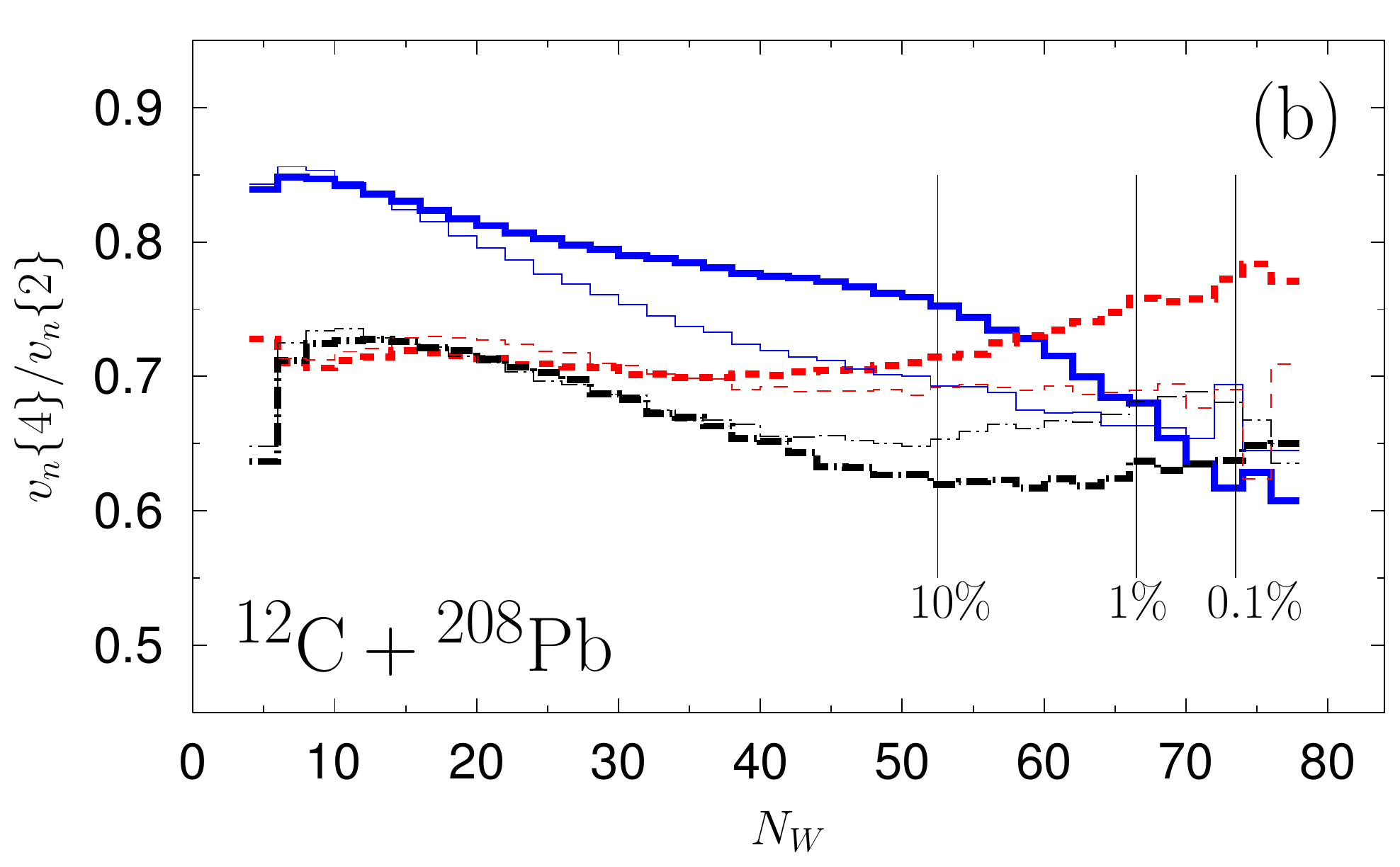}
%\vspace{-8mm}
\caption{(Color online) The same as in Fig.~\ref{fig:7Be} but for \nucCa\ collisions.}
\label{fig:12C}
\end{figure*}

\begin{figure*}[tb]
\centering
%\vspace{-2mm}
\includegraphics[angle=0,width=0.495 \textwidth]{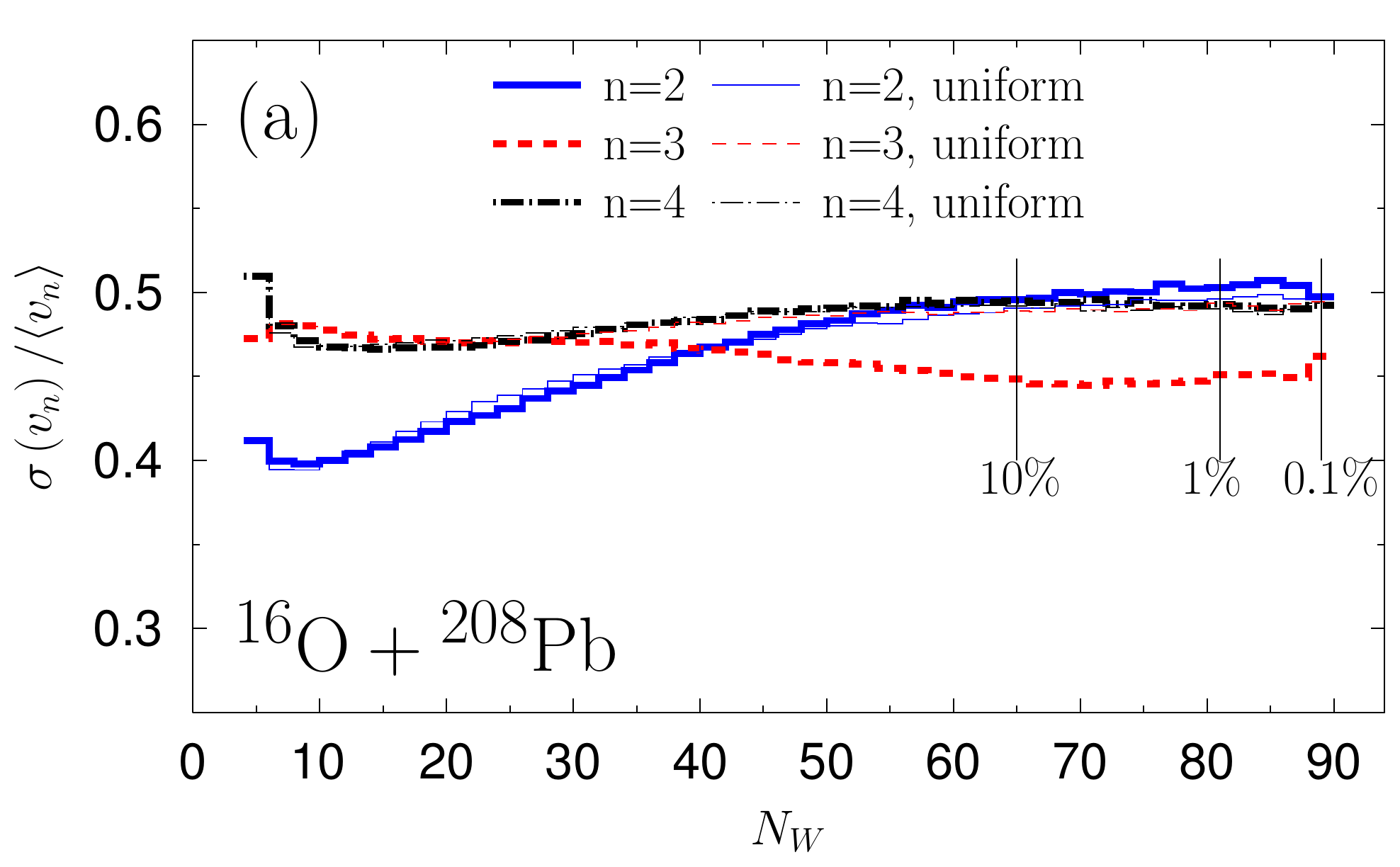}
\includegraphics[angle=0,width=0.495 \textwidth]{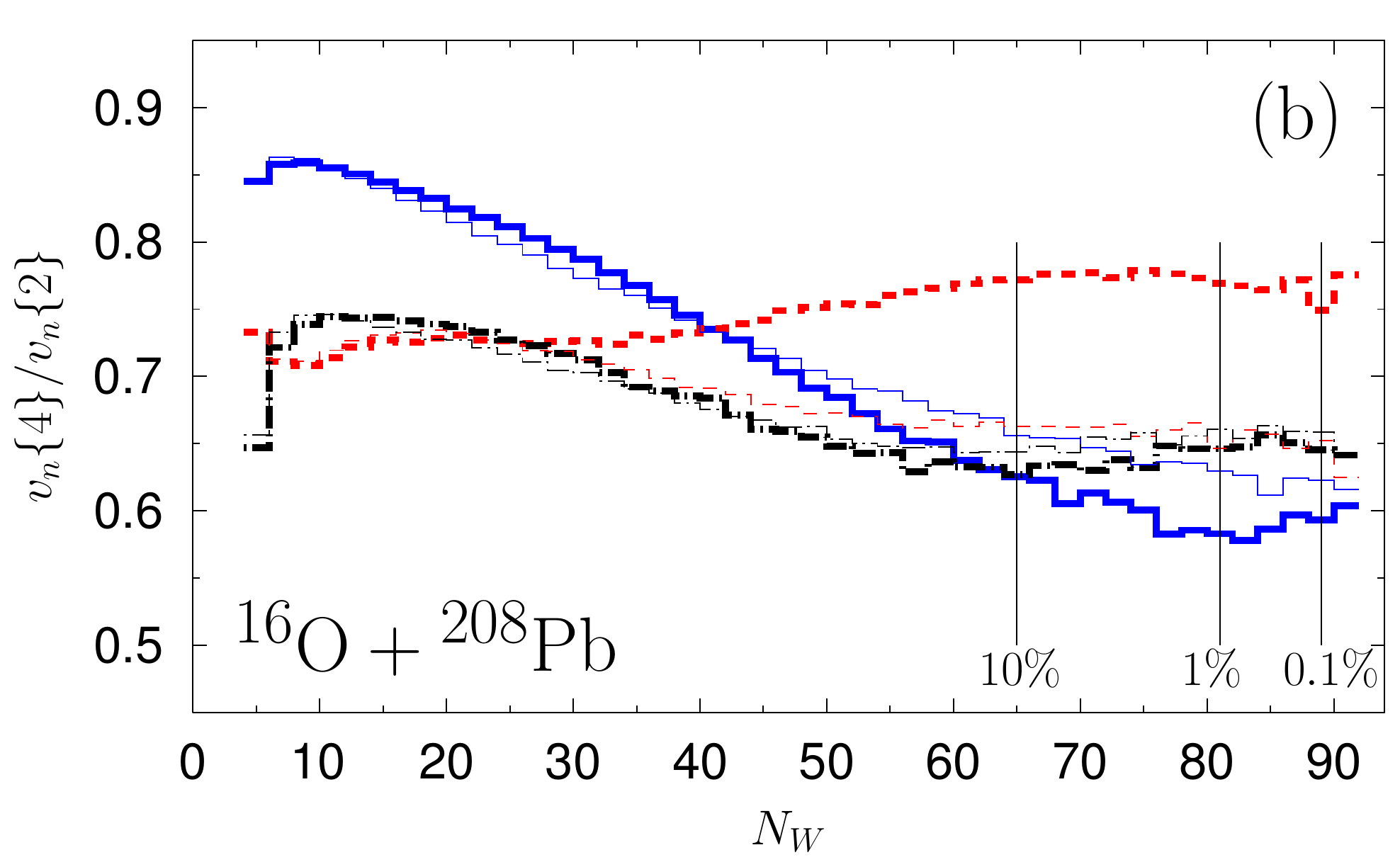}
%\vspace{-8mm}
\caption{(Color online) The same as in Fig.~\ref{fig:7Be} but for \nucOa\ collisions.}
\label{fig:16O}
\end{figure*}

\begin{figure*}[tb]
\centering
%\vspace{-2mm}
\includegraphics[angle=0,width=0.495 \textwidth]{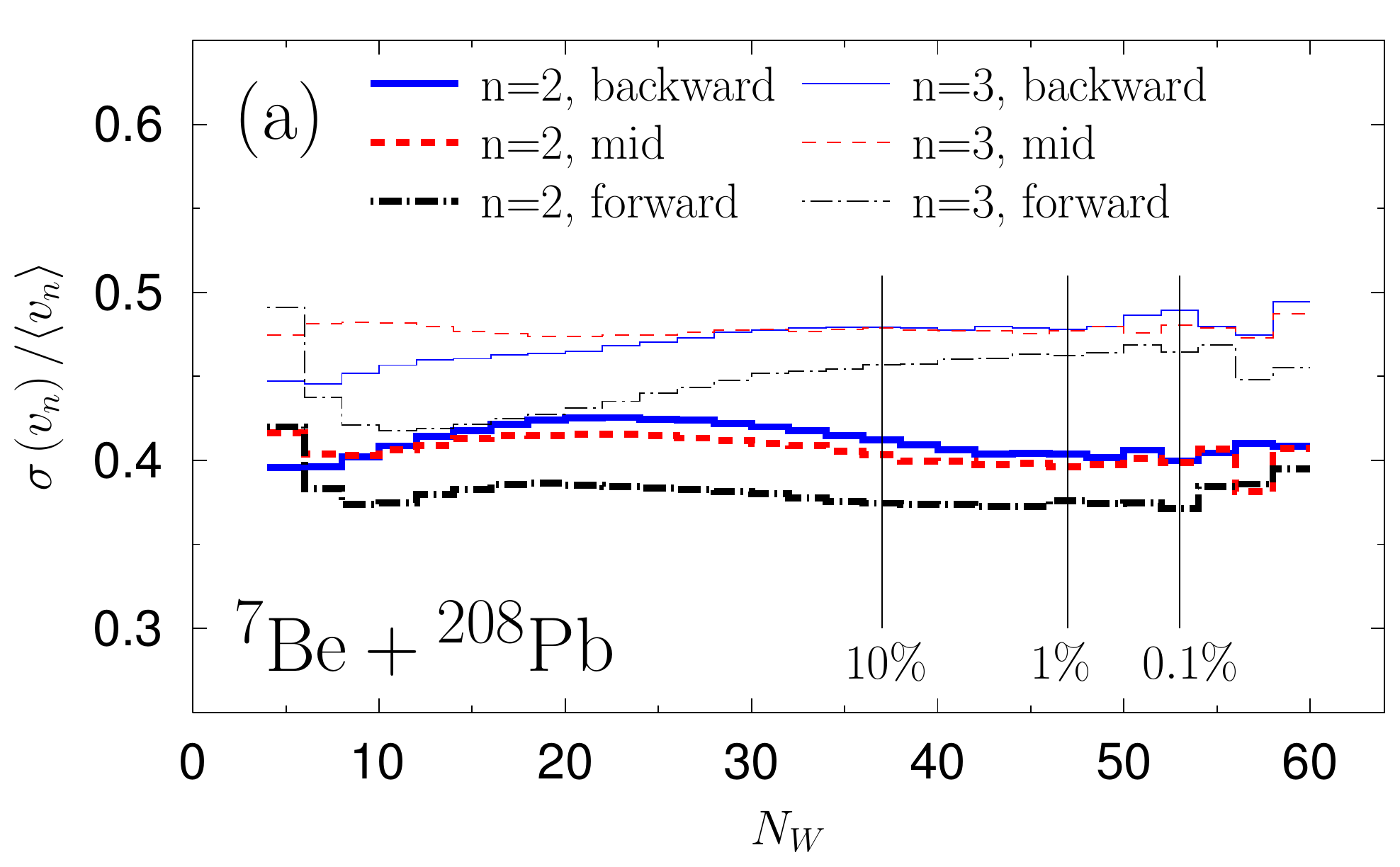}
\includegraphics[angle=0,width=0.495 \textwidth]{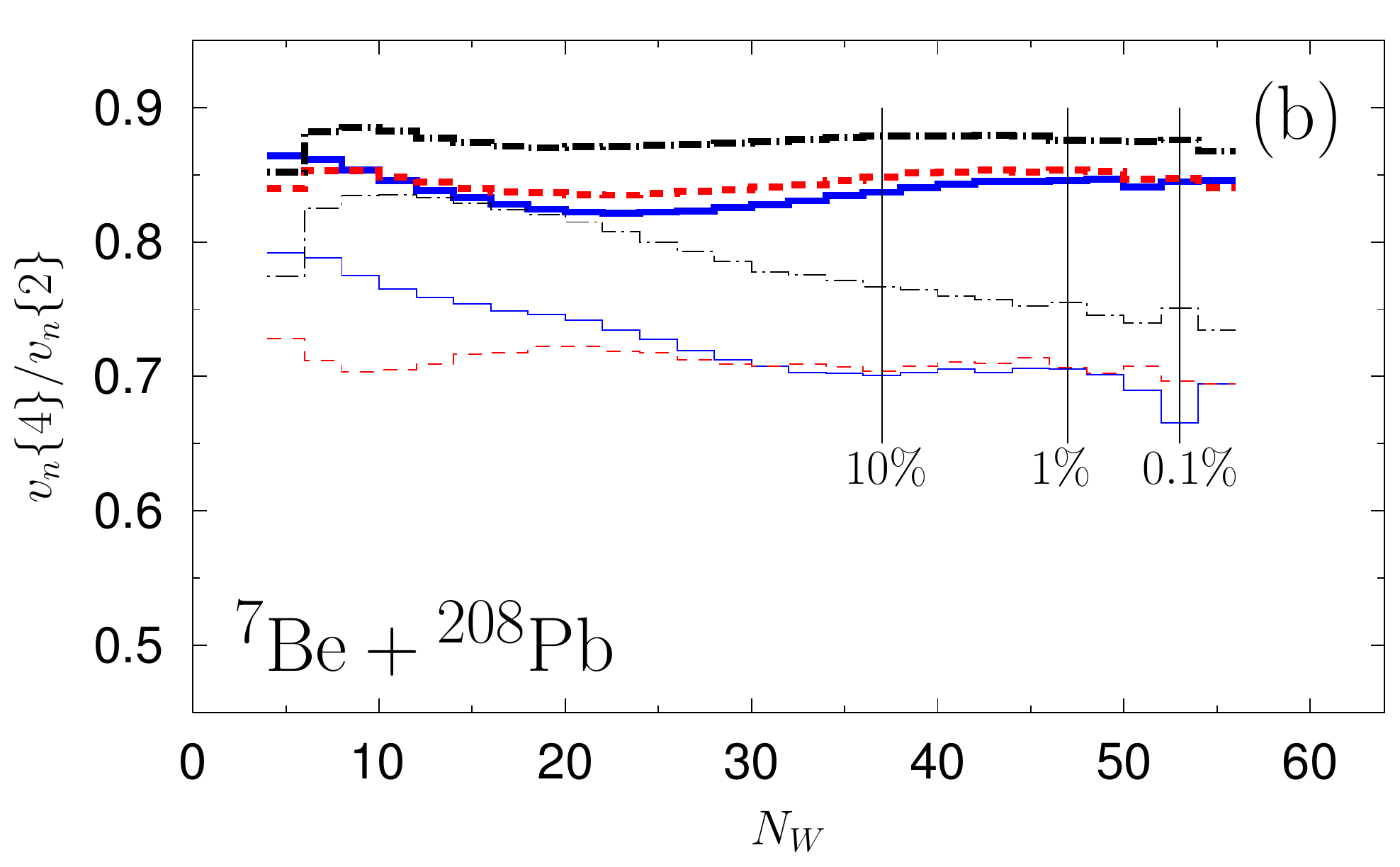}
%\vspace{-8mm}
\caption{(Color online) Scaled standard deviations of rank-$n$ flow coefficients (panel (a)) and ratios of the four-particle to two-particle 
cumulants (panel (b)) simulated for the backward, central, and forward rapidity regions, plotted as functions of the total number of the 
wounded nucleons. \nucBea\ collisions, clustered nuclei case. Thick lines correspond to the clustered case, and thin lines to the 
uniform distributions.}
\label{fig:7Berap}
\end{figure*}

To fix the parameters listed in Table~\ref{tab:param}, we use specific reference radial distribution obtained from 
VMC simulations, which use the Argonne~v18 two-nucleon and Urbana~X three-nucleon potentials, as provided in 
{\small \url{http://www.phy.anl.gov/theory/research/density}}~\cite{Wiringa:2013ala,Lonardoni:2017egu}. Our distribution parameters 
are then optimized such that the one particle densities $\rho(r)$ from VMC  
are properly reproduced.  Thus the radial density of the centers on nucleons serves as a constraint for building our clustered 
distributions. Figure~\ref{fig:density} shows the quality of our fit to the one-body densities, which is satisfactory in the 
context of modeling ultra-relativistic nuclear collisions.  We note from  Fig.~\ref{fig:density} that
the distributions (except for $^{7}$Be nucleus) develop a dip in the center.
The parameters used in our simulations are collected in Table~\ref{tab:param}.

As we are interested in specific effects of clusterization, as a ``null result'' we use the {\em uniform} distributions, i.e., 
with no  clusters.  We prepare such distributions with exactly the same radial density as the clustered ones. 
This is achieved easily with a trick, where we randomly re-generate the spherical angles of the nucleons from 
the clustered distributions, while leaving the radial coordinates intact.

\section{Harmonic flow in relativistic light-heavy collisions}
\label{sec:signatures}

As already mentioned in the Introduction, we use the so-called Glauber approach to model the early stage of the collision. The Glauber model~\cite{Glauber:1959aa}
formulated almost sixty years ago to model the elastic scattering amplitude in high-energy collisions, was later 
extended to inelastic collisions~\cite{Czyz:1968zop}, and subsequently led to the widely used wounded-nucleon model~\cite{Bialas:1976ed}.
The model assumes that 
the trajectories of nucleons are straight lines and the individual nucleons at impact parameter $b$ interact with a probability $P(b)$, where 
$\int d^2b \, P(b) =\sigma_{\rm inel}$ is the total inelastic nucleon-nucleon cross section. We use a Gaussian form of $P(b)$, 
which for the studied heavy-ion observables is of sufficient accuracy~\cite{Rybczynski:2011wv}.

Generally, in the Glauber framework, at the initial stage of the collision the interacting nucleons deposit 
entropy (or energy) in the transverse plane. Such deposition occurs from wounded nucleons, but also from binary collisions.
Such an admixture of binary collisions is necessary to obtain proper multiplicity distributions~\cite{Kharzeev:2000ph,Back:2001xy}.
In this model the transverse distribution of entropy takes the form 
\begin{eqnarray}
\rho(x,y)&=&\frac{1-\alpha}{2}\rho_W(x,y) + \alpha \rho_{\rm bin}(x,y), \label{eq:rho0}
\end{eqnarray}
where $\rho_W(x,y)$ is the distribution of the wounded nucleons, 
$\rho_{\rm bin}(x,y)$ is the distribution of the binary collisions, and  $\alpha$ is the parameter controlling the relative weight of the wounded to binary sources. 
In our simulations we use $\alpha=0.12$ (the value fitting the multiplicity distributions at the SPS collision energies). 
The sources forming the distributions are smeared with a Gaussian of a width of 0.4~fm.

In the following we show the numerical results of our GLISSANDO~\cite{Broniowski:2007nz,Rybczynski:2013yba} 
simulations of collisions of the described above nuclei composed 
of $\alpha$-clusters with ${}^{208}$Pb nucleus at $\sqrt{s_{NN}}=17$~GeV, where the corresponding inelastic nucleon-nucleon 
cross section is $\sigma_{\rm inel}=32$~mb. Such collision energies are available at SPS and the considered reactions are possible to study in the on-going 
NA61/SHINE experiment with ${}^{208}$Pb or proton beams. A variety of targets and secondary beams are available in this experiment~\cite{Abgrall:2014xwa}.
Therefore the present study may be thought of as a case study for possible NA61/SHINE investigations.

To analyze the effects of clusterization in the considered light nuclei on the harmonic flow coefficients
in the reactions with  ${}^{208}$Pb nuclei, one needs to use appropriate flow measures. The eccentricity coefficients, $\epsilon_n$, are designed
as measures of the harmonic deformation in the initial state. They are defined for each collision event as
\begin{eqnarray} 
\epsilon_n e^{i n \Phi_n} = - \frac{\int \rho(x,y) e^{i n\phi}  (x^2+y^2)^{n/2} dx dy}{\int \rho(x,y) (x^2+y^2)^{n/2}  dx dy}, \label{eq:eps}
\end{eqnarray}
for $n=2,3,\dots$, with $\phi=\arctan(y/x)$ and $\Phi_n$ denoting the angle of the principal axes in the transverse plane $(x,y)$.

The subsequent collective evolution with hydrodynamic~\cite{Heinz:2013th,Gale:2013da,Jeon:2016uym} 
or transport~\cite{Lin:2004en} has a shape-flow transmutation feature: The deformation of shape in the initial stage leads to harmonic 
flow of the hadrons produced in the late stage. The effect is manifest in an approximate proportionality of the flow 
coefficients $v_n$ to the eccentricities  $\epsilon_n$, which holds for $n=2$ and $3$ (for higher rank non-linear coupling effects may be present):
\begin{equation}
v_n = \kappa_n \epsilon_n \ , 
\label{eq:linear}
\end{equation} 
The cumulant coefficients follow an analogous relation: 
\begin{eqnarray}
v_n\{m\}=\kappa_n \epsilon_n\{m\}.
\end{eqnarray}
The proportionality coefficients $\kappa_n$ depend on various features of the colliding system (centrality, collision energy), 
but are to a good approximation independent of the eccentricity itself, hence the above relations are linear.
To get rid of the influence of the (generally) unknown $\kappa_n$ coefficients on the results, 
one may consider the ratios of cumulants of different order $m$ for a given rank-$n$ flow coefficient $v_n$, e.g., 
\begin{eqnarray}
\frac{v_n\{m\}}{v_n\{2\}}=
\frac{\epsilon_n\{m\}}{\epsilon_n\{2\}} \ . \label{eq:ratios} 
\end{eqnarray}
Therefore the ratios of the flow cumulants can be directly compared to the corresponding ratios of the eccentricity cumulants.
In our work we also use the scaled event-by event standard deviation, ${\sigma(\epsilon_n)}/{\langle \epsilon_n \rangle}$, where
\begin{eqnarray}
 \frac{\sigma(\epsilon_n)}{\langle \epsilon_n \rangle} \simeq \frac{\sigma(v_n)}{\langle v_n \rangle}. \label{eq:ev}
\end{eqnarray}

In order to find the specific effects of clusterization, we always compare the obtained results to those corresponding to the  ``uniform'' case, where 
the nucleons are distributed without clusterization (see Sect.~\ref{sec:making}).

In Figs.~\ref{fig:7Be} and~\ref{fig:9Be} we show the event-by-event scaled standard deviations of the elliptic ($n=2$), 
triangular ($n=3$), and quadrangular ($n=4$) flow coefficients, as well as the ratios of the four-particle to 
two-particle cumulants, plotted as functions of the total number of wounded nucleons. Since clusters in ${}^{7,9}$Be nuclei form a dumbbell 
shape, the influence of clusterization is, as expected, visible in the $n=2$ (elliptic) coefficients. 
The behavior seen in panels (a) is easy to explain qualitatively: At large numbers of wounded nucleons the beryllium is oriented in such a way that it 
hits the wall of ${}^{208}$Pb side-wise, as drawn in Fig.~\ref{fig:concept}. Then the eccentricity of the created fireball, which is an imprint of the 
intrinsic shape of beryllium, is largest. Hence the scaled variance decreases (note division with $\langle \epsilon_n \rangle$ in Eq.~(\ref{eq:ev})) with $N_w$.
The feature is clearly seen from Figs.~\ref{fig:7Be} and~\ref{fig:9Be}. Of course, this is not the case for the uniform distributions, where at large 
$N_w$ the scaled standard deviations for all $n$ acquire similar values. A detailed quantitative 
understanding of the dependence on $N_w$ requires simulations, as one needs to assess the influence of the random fluctuations on eccentricities, or account for effects 
when the beryllium hits the edge of ${}^{208}$Pb. The size of the effect in panels (a) starts to be significant for the 10\% of the highest-multiplicity events.

The results for the $v_n\{4\}/v_n\{2\}$ (panels (b) of Figs.~\ref{fig:7Be} and~\ref{fig:9Be}) are complementary.
We note that for high multiplicity collisions
the ratio is significantly larger for the clustered case compared to the uniform distributions. This is because the 
two-particle cumulants are more sensitive to the random fluctuations than the four-particle cumulants.

For the case of ${}^{12}$C+${}^{208}$Pb and ${}^{16}$O+${}^{208}$Pb collisions, the significant influence of clusters 
as compared to ``uniform'' case is visible for the rank-3 (triangular) coefficients, see Figs.~\ref{fig:12C} and \ref{fig:16O}. 
This is mainly caused by the triangular and tetrahedral arrangements of clusters in ${}^{12}$C and ${}^{16}$O, respectively.
The qualitative understanding is as for the beryllium case, with the replacement of $n=2$ with $n=3$. The case of ${}^{12}$C 
has also been thoroughly discussed in Ref.~\cite{Bozek:2014cva}.

All previously shown simulations were carried out at the mid-rapidity, $y\sim 0$, region. To study the dependence on rapidity, 
we apply a model with rapidity-dependent emission functions of the entropy sources. Such an approach is necessary,
since in most fixed-target experiments the detectors measure particles produced in rapidity regions which are away from the mid-rapidity
domain. 
Taking this into account, we apply the model described in Refs.~\cite{Bialas:2004su,Bozek:2010bi}. There, the initial density of the fireball 
in the space-time rapidity $\eta_\parallel=\frac{1}{2} \log (t+z)(t-z)$ and the transverse coordinates ($x, y$) is described by the function:
\begin{eqnarray}
\rho(\eta_\parallel,x,y)&=&(1-\alpha)[\rho_A(x,y) f_+(\eta_\parallel)
+ \rho_B(x,y) f_-(\eta_\parallel)] \nonumber \\
&+& \alpha
\rho_{\rm bin}(x,y) \left [ f_+(\eta_\parallel) + f_-(\eta_\parallel) \right ].
\label{eq:em}
\end{eqnarray}
which straightforwardly generalizes Eq.~(\ref{eq:rho0}), assuming factorized profiles from a given source.
Here $\rho_{A,B}(x,y)$ denotes the transverse density of the wounded sources from the nuclei $A$ and $B$, 
which move in the forward and backward directions, respectively. The entropy emission functions  
$f_{\pm}(\eta_\parallel)$ are given explicitly in~\cite{Bozek:2010bi}. They are peaked in the forward or backward 
directions, respectively, reflecting the fact that a wounded nucleon emits preferentially in its own forward hemisphere.

In the Fig.~\ref{fig:7Berap} we plot, as functions of $N_w$, the scaled standard deviations of the rank-2 and 3 flow coefficients and ratios 
of the four-particle to two-particle cumulants calculated in backward ($\eta_\parallel = -2.5 $), central ($\eta_\parallel = 0$), and forward ($\eta_\parallel = 2.5 $) 
rapidity regions (at the SPS collision energy of $\sqrt{s_{NN}}=17$~GeV the rapidity of the beam is $\sim 2.9$). 
We focus on results here for \nucBea\ collisions, as for the other light clustered nuclei the results
are qualitatively similar. The centrality dependence 
of the scaled standard deviation of second-rank flow coefficients (panel (a)) is similar for all considered regions of phase-space, however its magnitude 
grows when we move from the backward (${}^{208}$Pb) to the forward (beryllium) hemisphere. The effect has to do with a 
a much larger number of wounded nucleons in the backward compared to the forward hemisphere in the applied model 
of the rapidity dependence. This makes the random fluctuations smaller in the backward compared to the forward hemisphere, giving the effect seen in 
Fig.~\ref{fig:7Berap}.

We note that the previously discussed difference of the behavior of eccentricities between the clustered and uniform cases holds also 
for other regions in rapidity, which makes the effect possible to study also in the fixed-target experiments with detectors covering the 
forward rapidity region.

\section{Proton-polarized light nucleus scattering}
\label{sec:pA}

In this section we present a more exploratory study, as the investigation needs the magnetic field 
to polarize the beryllium nuclei along a chosen direction. Polarized nuclear targets or beams have not yet been
used in ultra-relativistic collisions. Nevertheless, our novel effect, also geometric in its origin, is worth 
presenting as a possibility for future experiments.

Since the ground states of ${}^{7,9}$Be nuclei have $J^P=3/2^-$, they can be polarized. Then, due to their
cluster nature, the intrinsic symmetry axis correlated to the polarization axis in a specific way described in detail below. 
One can thus control (to a certain degree)  the orientation of the intrinsic dumbbell shape. This, in turn, can be probed in 
ultra-relativistic collisions with protons, as more particles are produced when the proton goes along the dumbbell, compared to the case when 
it collides perpendicular to the symmetry axis. 

We wish to consider the beryllium nuclei polarized in magnetic field, therefore the first task is to obtain 
states of good quantum numbers in our model approach, where we prepare intrinsic states 
with the method described in Sect.~\ref{sec:making}. 
We use the Peierls-Yoccoz projection (see, e.g., ~\cite{ring}), which is a 
standard tool in nuclear physics of (heavy) deformed nuclei. The basic formula to pass from an intrinsic wave function $\Psi^{\rm intr}_k(\Omega)$,
where $\Omega$ is the spherical angle of the symmetry axis and $k$ is the intrinsic spin projection, to the state of good quantum numbers $|j,m\rangle$ has the form
\begin{eqnarray}
|j,m\rangle = \sum_k \int d\Omega D^{j}_{m,k}(\Omega) |\Psi_k^{\rm intr}(\Omega) \rangle, \label{eq:PY}
\end{eqnarray}
where $ D^{j}_{m,k}(\Omega)$ is the Wigner $D$ function.

The ${}^7$Be nucleus has the following cluster decomposition and angular momentum decomposition between the spin of the clusters and the orbital angular momentum of the clusters:
\begin{eqnarray}
{}^7{\rm Be} &=& {}^4{\rm He} + {}^3{\rm He}, \label{eq:7Be} \\  
\frac{3}{2}^-  &=& 0^+ + \tfrac{1}{2}^+ + 1^-, \nonumber
\end{eqnarray}
where $0^+$ is the $J^P$ of the $\alpha$ particle,  $ \tfrac{1}{2}^+$ of ${}^3{\rm He}$, and $ 1^-$ is the orbital angular 
momentum. Similarly, for ${}^9$Be 
\begin{eqnarray}
{}^9{\rm Be} &=& {}^4{\rm He} + {}^4{\rm He} + n, \label{eq:9Be} \\  
\frac{3}{2}^-  &=& 0^+ + 0^+ +  \tfrac{1}{2}^+ + 1^-, \nonumber
\end{eqnarray}
where the neutron is assumed to be in an $S$ state, and the $J^P$ of the angular motion of the two $\alpha$ clusters is $1^-$. The Clebsch-Gordan decomposition is 
\begin{eqnarray}
|\tfrac{3}{2},m &=& \tfrac{3}{2}\rangle = |\tfrac{1}{2},\tfrac{1}{2}\rangle \otimes |1,1\rangle, \label{eq:CG} \\
|\tfrac{3}{2},m &=& \tfrac{1}{2}\rangle = 
\sqrt{\tfrac{2}{3}}|\tfrac{1}{2},\tfrac{1}{2}\rangle \otimes |1,0\rangle + \sqrt{\tfrac{1}{3}} |\tfrac{1}{2},-\tfrac{1}{2}\rangle \otimes |1,1\rangle. \nonumber 
\end{eqnarray}
In the intrinsic frame, where the clusters are at rest, the angular momentum comes from the spin of $ {}^3{\rm He}$ or $n$ in the cases of  ${}^7$Be 
or ${}^9$Be, respectively, hence the available values of $k$ are $\pm \tfrac{1}{2}$.

\begin{figure}
\begin{center}
\includegraphics[angle=0,width=0.4 \textwidth]{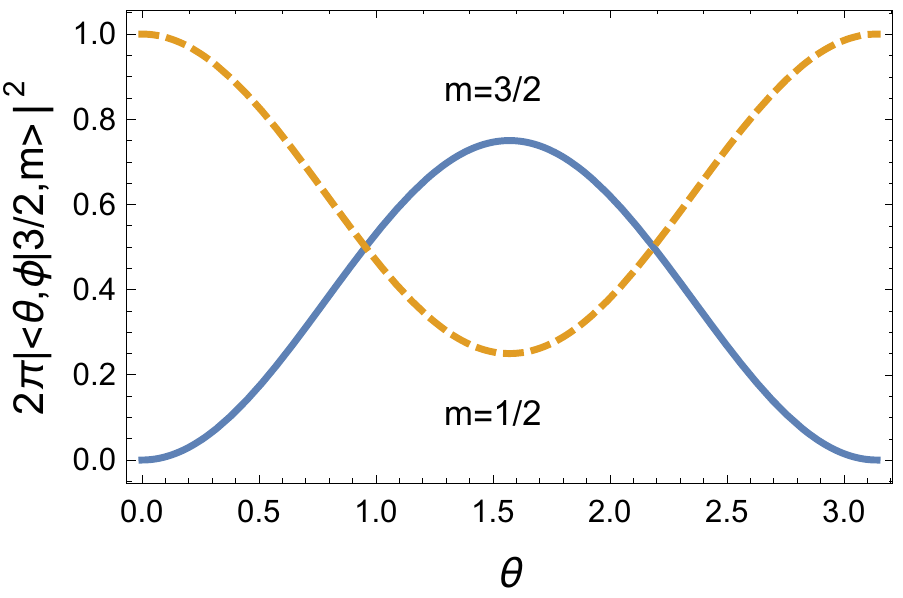} 
\caption{The distributions of the intrinsic symmetry axis of ${}^{7,9}$Be in the polar angle $\theta$, Eq.~(\ref{eq:tilt}), 
following from the Peierls-Yoccoz projection method. \label{fig:wig}}
\end{center}
\end{figure}

\begin{figure}
\includegraphics[angle=0,width=0.23 \textwidth]{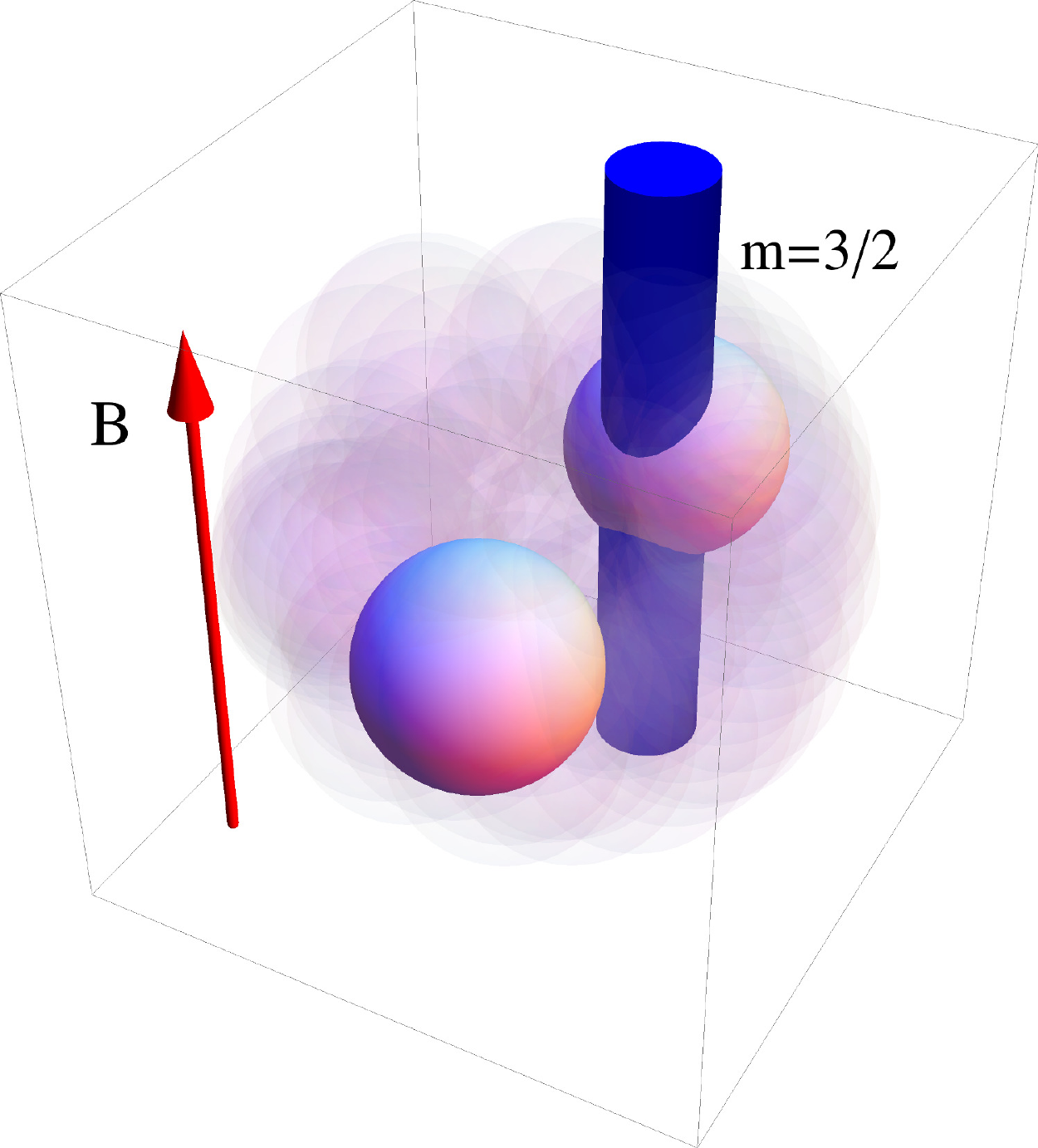} \includegraphics[angle=0,width=0.23 \textwidth]{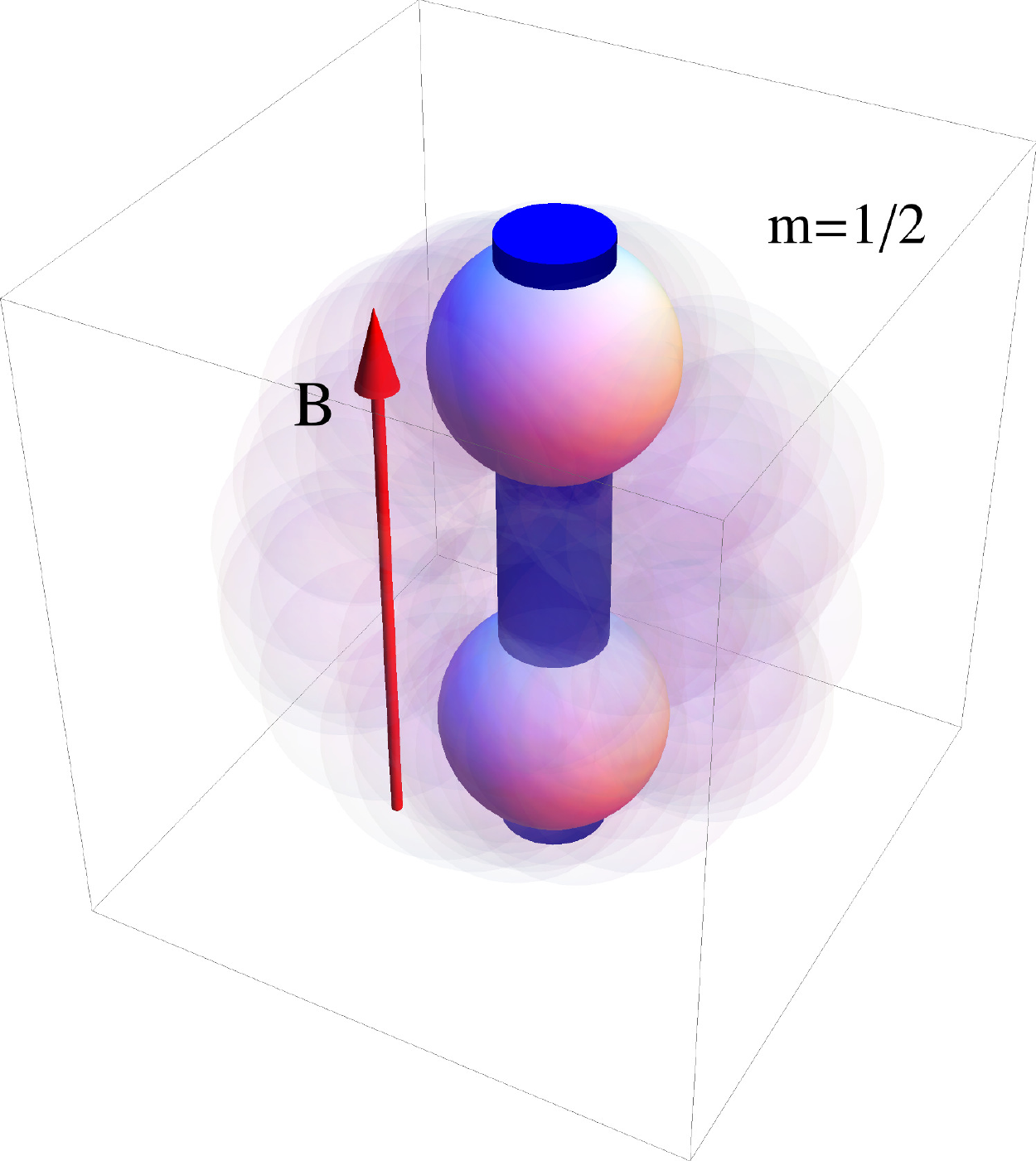} \\
\vspace{-2mm}
\includegraphics[angle=0,width=0.23 \textwidth]{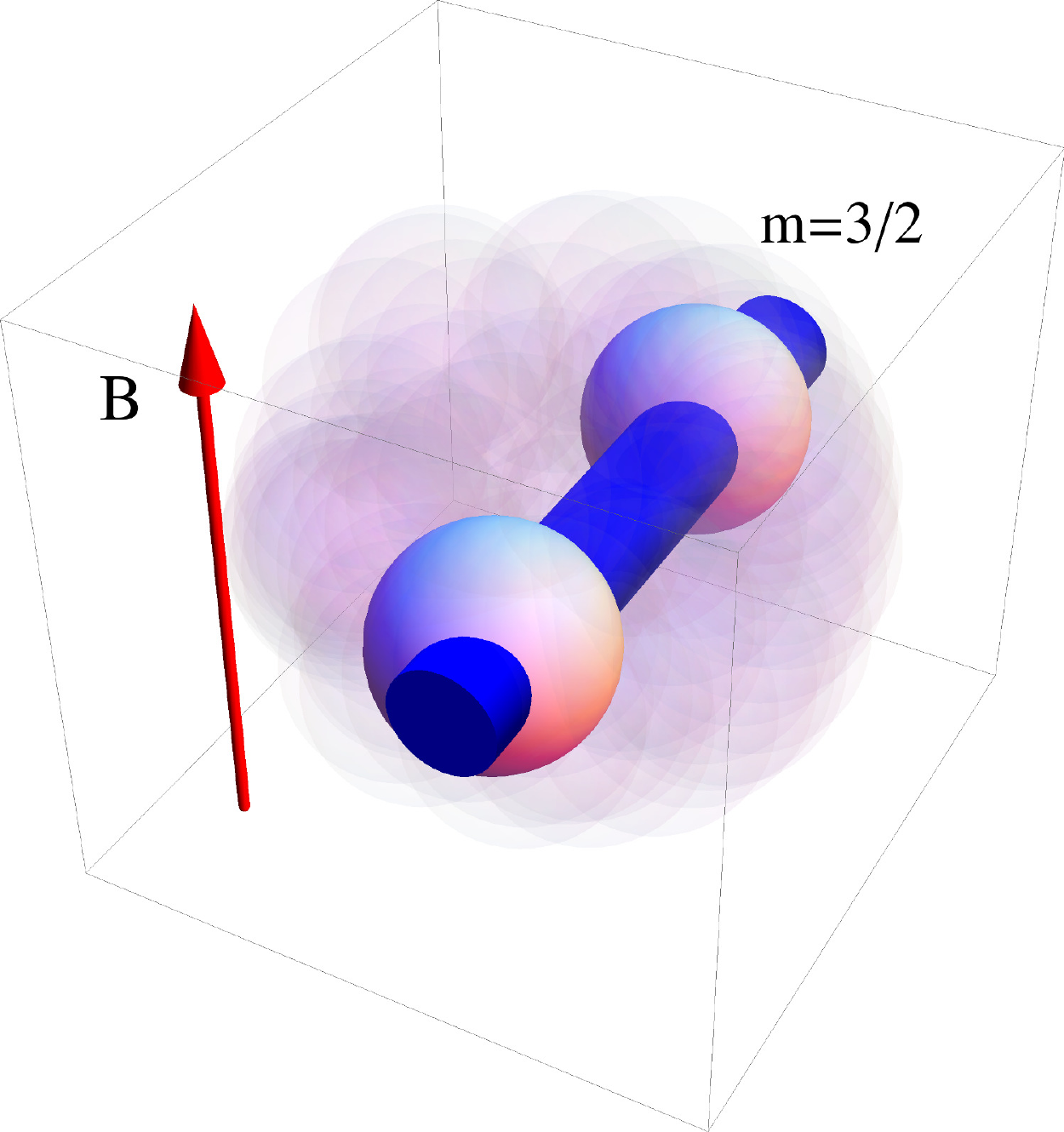} \includegraphics[angle=0,width=0.23 \textwidth]{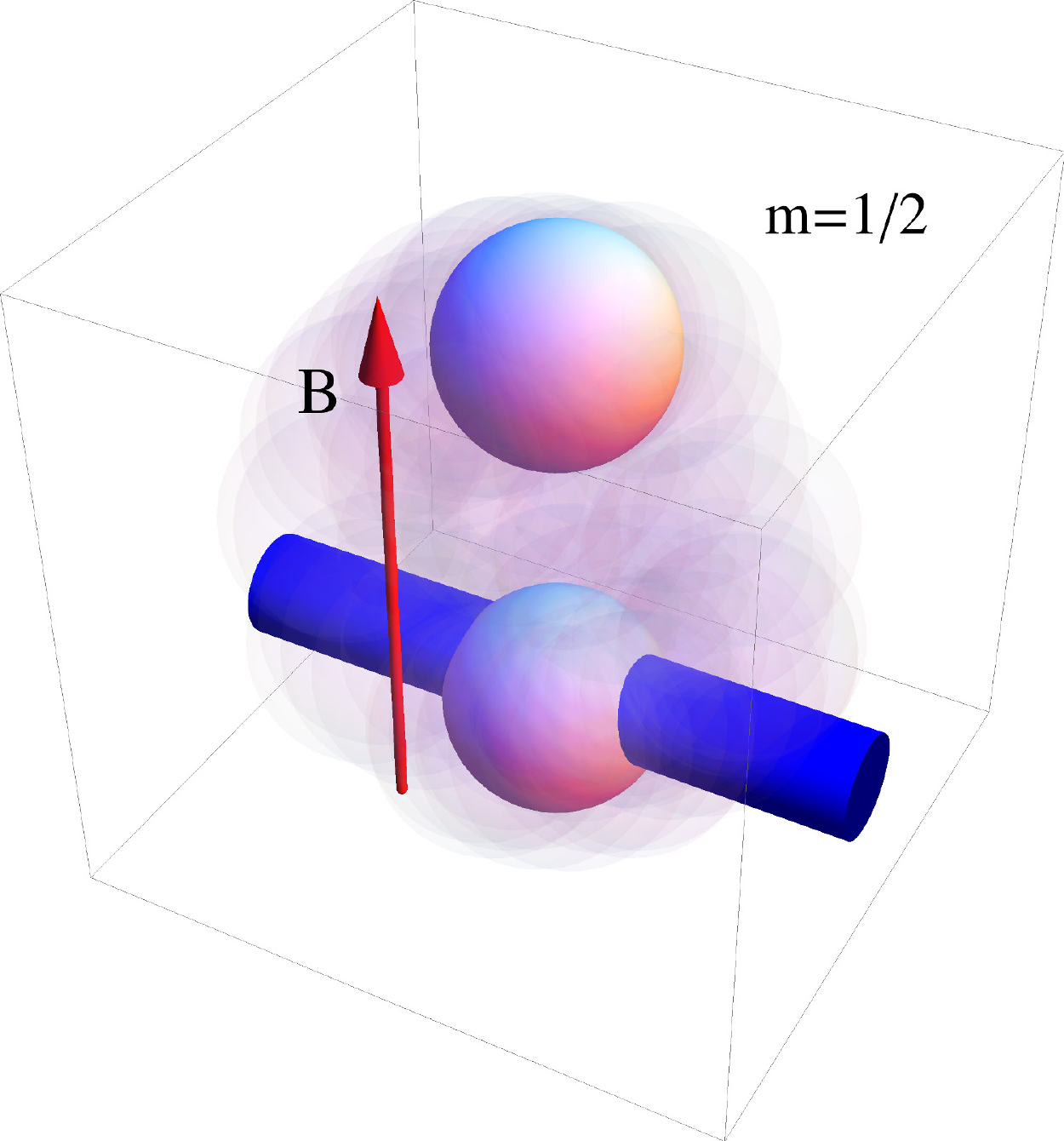} 
\vspace{-5mm}
\caption{Schematic representation of collisions of protons with polarized $^{7,9}$Be. The sphere presents the $\alpha$ or $^3$He clusters 
and the clouds indicate the quantum washing out of the symmetry axis of the intrinsic states, in accordance to Eq.~(\ref{eq:PY}). The tube represents 
the proton beam with the area given by the total inelastic proton-proton cross section. Arrows show the direction of the magnetic field which corresponds 
to the quantization axis of spin. Details in the text. \label{fig:geom}}
\end{figure}

\begin{figure*}
\includegraphics[angle=0,width=0.4 \textwidth]{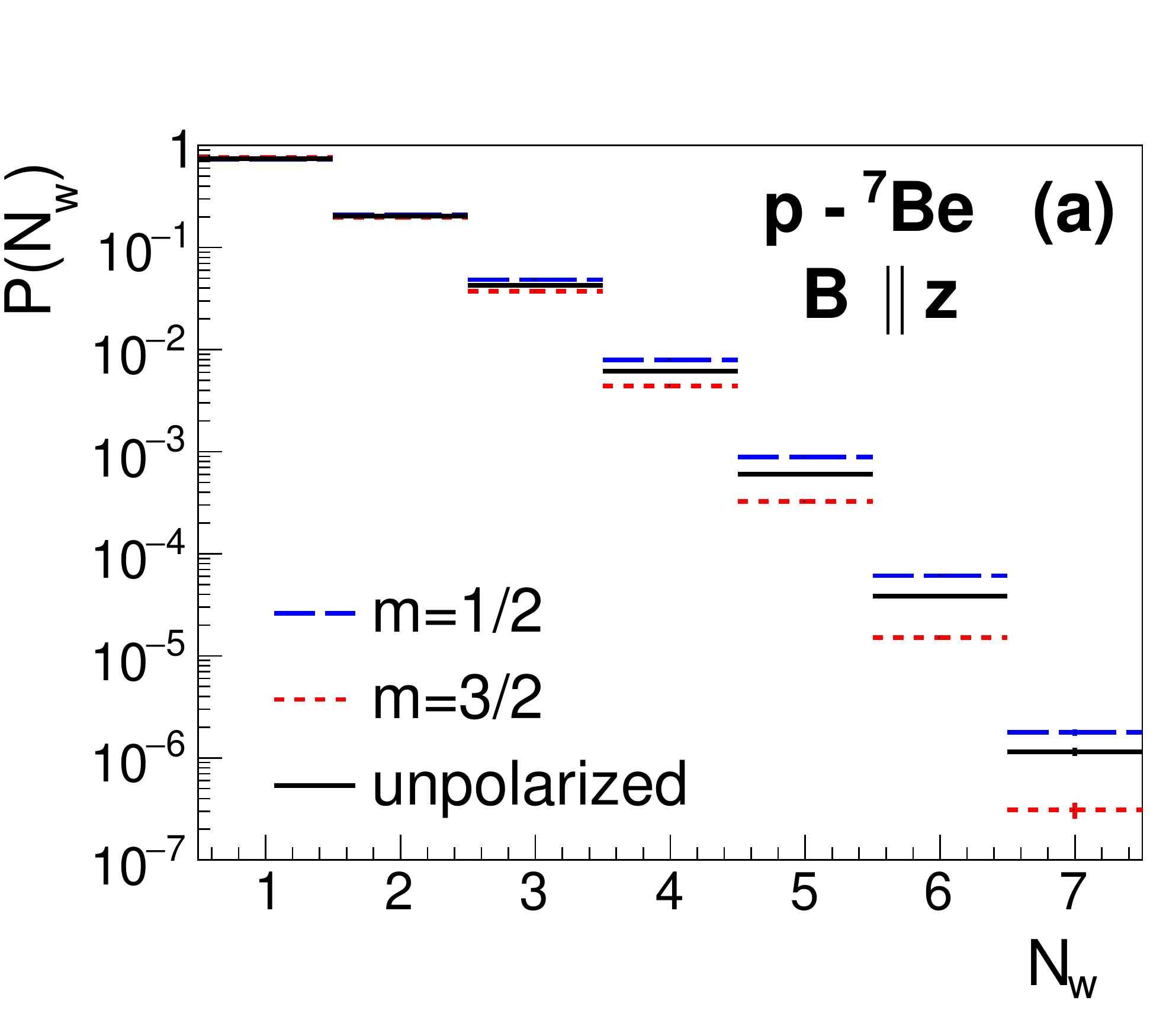} ~~ \includegraphics[angle=0,width=0.4 \textwidth]{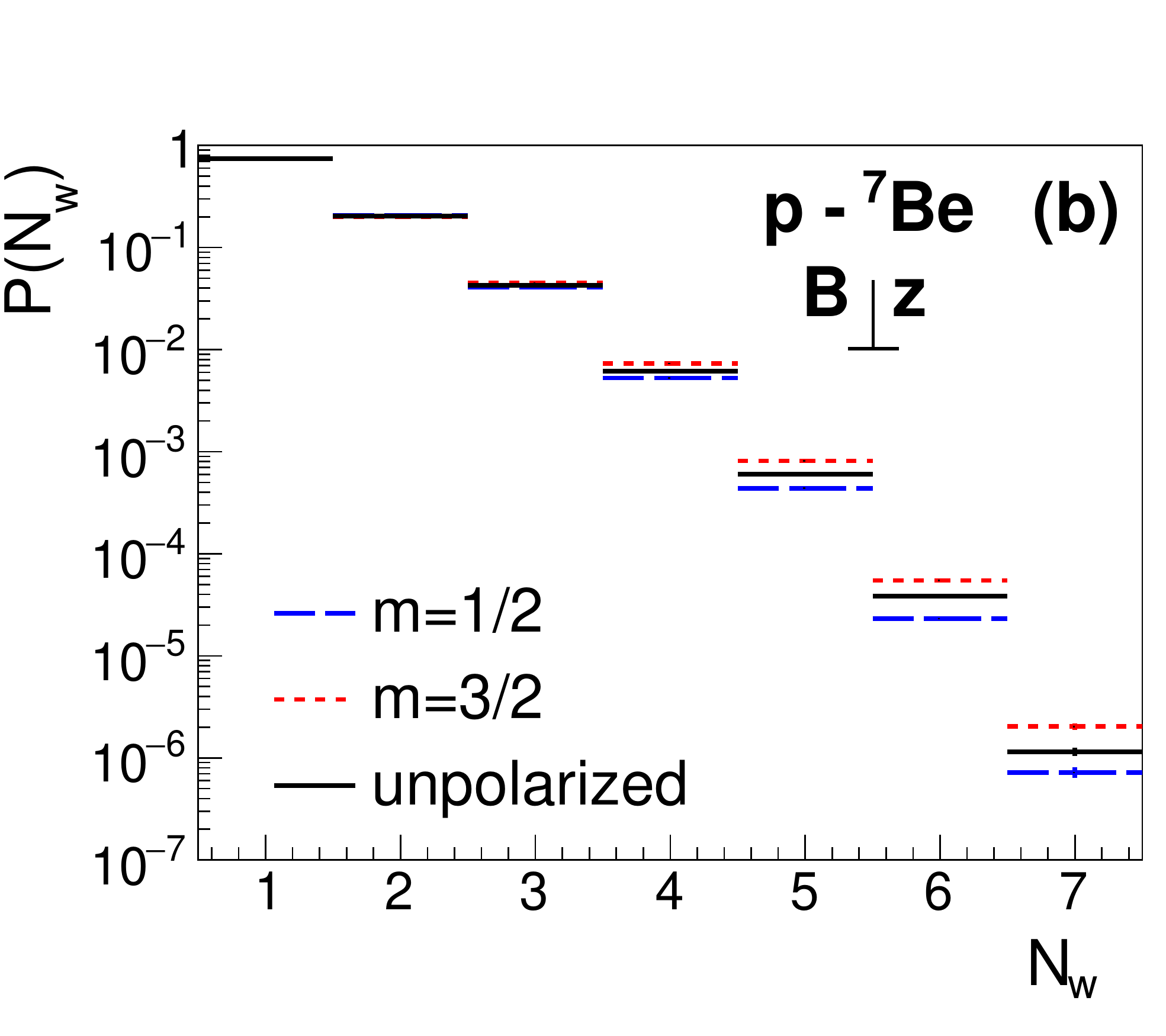}
\caption{Results of Monte Carlo simulations of p+$^7$Be collisions. We note that for $N_w \ge 3 $  the probability of wounding 
$N_w$ nucleons is higher for $m=1/2$ than for $m=3/2$ in the case when $\vec{B}$ is parallel to z axis (panel (a)). For situation when $\vec{B}$ 
is perpendicular to z, we observed more wounded nucleons for $m=3/2$ than for $m=1/2$ (panel (b)).   \label{fig:distr1}}
\end{figure*}
\begin{figure*}
\includegraphics[angle=0,width=0.4 \textwidth]{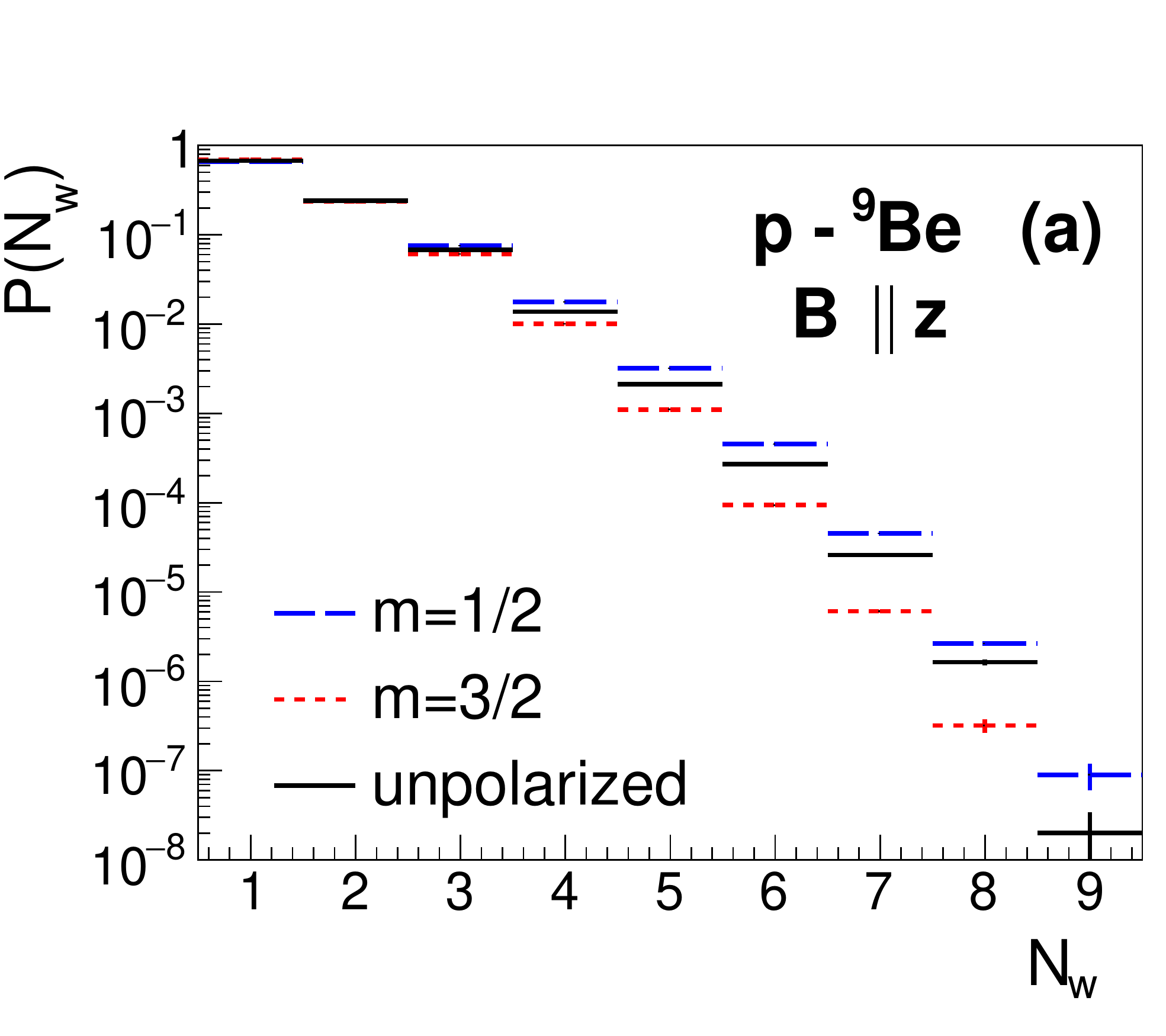} ~~ \includegraphics[angle=0,width=0.4 \textwidth]{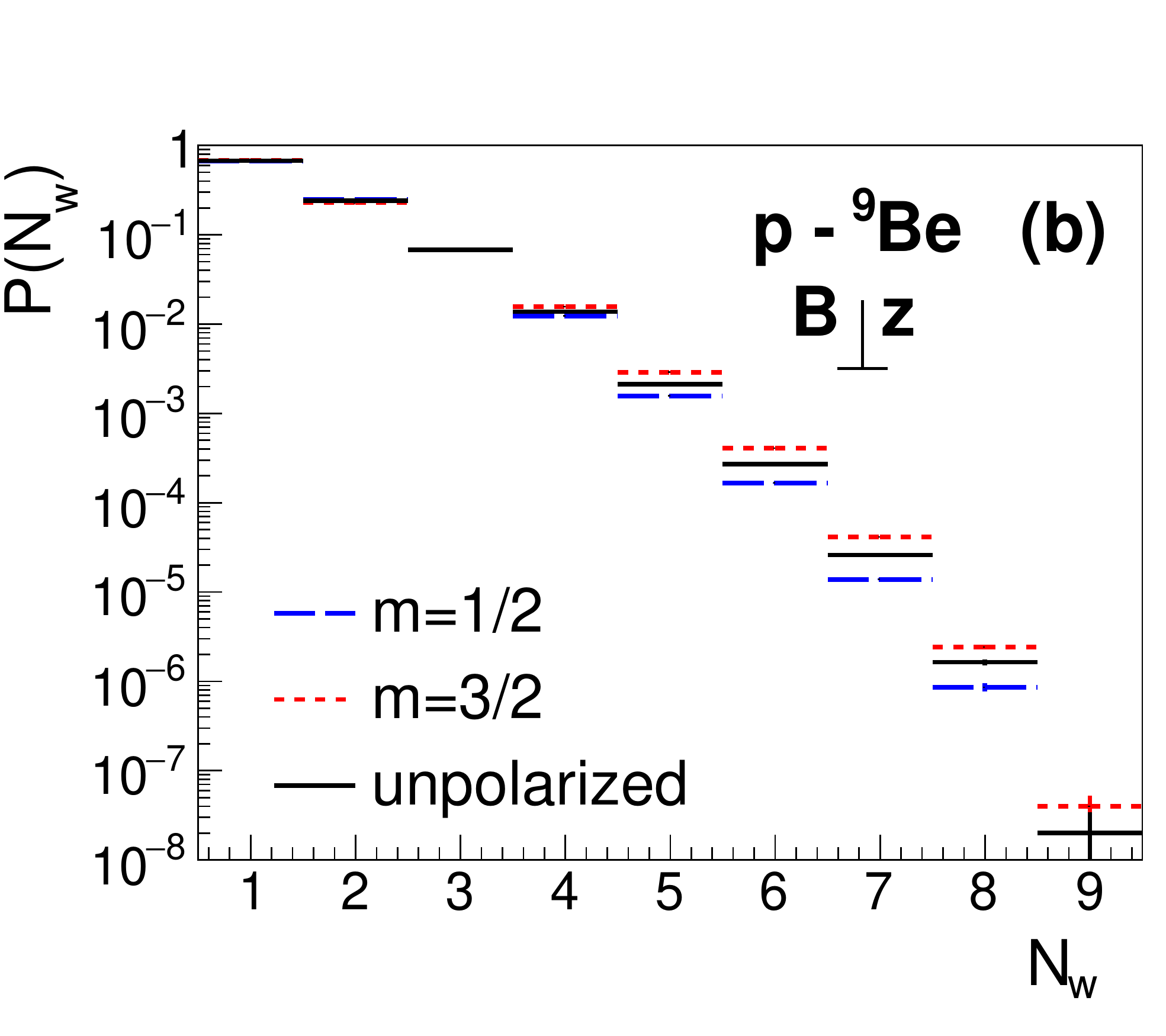} 
\caption{The same as in Fig.~\ref{fig:distr1} but for p+$^9$Be collisions. \label{fig:distr2}}
\end{figure*}

According to Eq.~(\ref{eq:PY}), we have for both nuclei 
\begin{eqnarray}
|\tfrac{3}{2},m\rangle = \sum_{k=\pm \tfrac{1}{2}} \int d\Omega D^{3/2}_{m,k}(\Omega) |\Psi_k^{\rm intr}(\Omega) \rangle. \label{eq:PY32}
\end{eqnarray}
Under the assumptions $\langle \Psi_{\rm intr}(\Omega') | \Psi_{\rm intr}(\Omega) \rangle \simeq \delta(\Omega-\Omega')$, which becomes 
exact in the limit of many nucleons, but still holds to a sufficiently good accuracy for 7 or 9 nucleons, we find 
\begin{eqnarray}
\left | \langle \theta, \phi |\tfrac{3}{2},m\rangle \right |^2 =  [D^{3/2}_{m,1/2}(\theta,\phi)]^2+[D^{3/2}_{m,-1/2}(\theta,\phi)]^2. 
\end{eqnarray}
Explicitly, 
\begin{eqnarray}
&& \left | \langle \theta, \phi |\tfrac{3}{2},\tfrac{3}{2} \rangle \right |^2 = |Y_{11}(\theta,\phi)|^2= \frac{3}{8\pi} \sin^2 \theta, \label{eq:tilt} \\
&& \left | \langle \theta, \phi |\tfrac{3}{2},\tfrac{1}{2} \rangle \right |^2 = \tfrac{2}{3} |Y_{10}(\theta,\phi)|^2 + \tfrac{1}{3} |Y_{11}(\theta,\phi)|^2 \nonumber \\
&& \hspace{2cm} =  \frac{1}{8\pi} \left ( 1+3\cos^2 \theta \right ), \nonumber
\end{eqnarray}
in accordance to Eq.~(\ref{eq:CG}). The distributions (\ref{eq:tilt}), which depend on the polar angle $\theta$ and 
not on the azimuthal angle $\phi$, are shown in Fig.~\ref{fig:wig}.

The prescription for the Monte Carlo simulations that follows from the above derivation is that the symmetry axes of ${}^{7,9}$Be should be 
randomly tilted in each collision event according to the distributions (\ref{eq:tilt}). We note that the $m=1/2$ state is approximately aligned 
along the spin projection axis (the distribution peaks at $\theta=0$ or $\theta= \pi$), 
whereas the $m=3/2$ state is distributed near the equatorial plane (with the maximum at $\theta=\pi/2$). 

Suppose that the targets of ${}^{7,9}$Be are 100\% polarized along the direction of the magnetic field $B$ and consider collisions with a proton beam 
parallel or perpendicular to $B$. Then the geometry of the collision is influenced by the distributions of the intrinsic symmetry axis, 
as pictorially displayed in Fig.~\ref{fig:geom}.
 
The figure shows schematically the collisions of protons with a polarized $^{7,9}$Be target, with the spheres representing the $\alpha$ or $^3$He 
clusters and the clouds indicating the quantum distribution of the symmetry axis of the intrinsic states, in accordance to Eq.~(\ref{eq:PY}). In the two 
left panels of the Fig.~\ref{fig:geom}, corresponding to $m=3/2$ states the clusters are distributed near the equatorial plane, whereas in the two right 
panels, corresponding to $m=1/2$ states the distribution of the clusters is approximately align along the quantization axis given by the magnetic field 
direction B. The tubes represent the proton beam, drawn in such a way that the area of the tube is given by the total inelastic proton-proton cross section.
We can distinguish several geometric cases in the top panels of Fig.~\ref{fig:geom} the proton beam is parallel to the direction of $\vec{B}$, we notice 
that for the $m=1/2$ case the chance of hitting two clusters, thus wounding more nucleons, is higher than for $m=3/2$ case. The effect is opposite 
when the proton beam is perpendicular to $\vec{B}$ as can be seen from the two bottom panels. 

The above discussed simple geometric mechanism finds its realization in numerical Monte Carlo simulations. Distribution of the number of 
wounded nucleons (in a logarithmic scale) is shown in Fig.~\ref{fig:distr1} and in Fig.~\ref{fig:distr2}. We note from panels (a) that in the 
case of $\vec{B}$ parallel to z (beam direction) indeed the probability of wounding more nucleons $N_w\ge 3$ is larger for $m=1/2$ than 
for $m=3/2$. The effect for $N_w=5$ reaches about a factor of 5, and increases for higher $N_w$. Note however, that at higher $N_w$ the collisions 
become very rare, thus statistical errors would preclude measurements.
In the case when $\vec{B}$ is perpendicular to z (panels (b)) the effect is opposite 
with higher probability of wounding more nucleons for $m=3/2$ than for $m=1/2$.

\section{Summary and conclusions}
\label{sec:summa}

We have shown that clusterization in light nuclei leads to characteristic signatures which could be studied in ultra-relativistic nuclear collisions. 
The presence of clusters leads to specific intrinsic geometric deformation, which in collisions with a heavy nucleus 
generates hallmark harmonic flow patterns, especially 
for the  collisions of highest multiplicity of the produced particles. 
As the phenomenology of flow and the corresponding data analysis methods are standard, we believe that the proposal 
is experimentally feasible, requiring collisions with appropriate beams and then using the well developed and tested data analysis 
techniques. We note that in the NA61/SHINE experiment the beryllium beams and targets, studied in this paper, 
have already been used~\cite{Abgrall:2014xwa}. 

We have also explored an opportunity following from the fact that the ground states of  ${}^{7,9}$Be have a non-zero spin, which allows for their 
polarization in an external magnetic field. Then, clusterization leads to significant effects in the spectra of participant (or spectator) nucleons in ultra-relativistic 
collisions with the protons. We have found a factor of two effects for $N_w = 4 $ and an order of magnitude effect for $N_w \ge 6$, when changing the orientation of the direction of the beam 
relative to the polarization axis, or when comparing the spin states \mbox{$m=3/2$} and \mbox{$m=1/2$}. As the polarized nuclei have not, 
up to now, been used in ultra-relativistic nuclear collisions, our proposal is to be considered in future experimental proposals.

Finally, we note that the effects of $\alpha$ clusterization for heavier nuclei are small in the sense that the resulting intrinsic eccentricities are much 
smaller than in the light systems considered in this paper. Therefore the investigations with the ${}^{7,9}$Be, ${}^{12}$C, and ${}^{16}$O nuclei 
would be most promising. 

\begin{acknowledgments}
The numerical simulations were carried out in laboratories created under the project
``Development of research base of specialized laboratories of public universities in Swietokrzyskie region'',
POIG 02.2.00-26-023/08, 19 May 2009.
MR was supported by the Polish National Science Centre (NCN) grant 2016/23/B/ST2/00692 and WB by NCN grant 2015/19/B/ST2/00937.
\end{acknowledgments}

\bibliography{clusters,hydr,from_hep}

\end{document}